\documentclass[a4paper,12pt]{article}

\usepackage{amsmath,amssymb,amsfonts} % Typical maths resource packages
\usepackage{graphicx}      % Packages to allow inclusion of graphics
\usepackage{color}                   % For creating coloured text and background
\usepackage{hyperref}
\usepackage{cite}
\definecolor{red}{rgb}{1,0,0}           % Standard colours red, green, blue
\definecolor{green}{rgb}{0,1,0}
\definecolor{blue}{rgb}{0,0,1}
\definecolor{darkblue}{rgb}{0,0,0.5}
\definecolor{lightblue}{rgb}{.5,.5,1}
\definecolor{lightgray}{gray}{.87}          % How you can define your own greys
\definecolor{Dark}{gray}{.20}
\definecolor{pink}{rgb}{.95,0.82,0.92}  % How you can define your own colours
\definecolor{yellow}{rgb}{1,1,0}
\definecolor{lightyellow}{rgb}{1,1,.5}
\definecolor{purple}{rgb}{0.7,0,0.85}
\definecolor{darkgreen}{rgb}{0,0.5,0}
\definecolor{orange}{rgb}{0.8,0.2,0.2}
\def \be {\begin{equation}}
\def \ee {\end{equation}}
\def \bea {\begin{eqnarray}}
\def \eea {\end{eqnarray}}
\def \nn {\nonumber}

\def \rr {\raise.35ex\hbox{\small $\prime$}\kern-.17em{\mbox{\large $\imath$}}}

\def \dels {\partial\kern-.5em / \kern.5em}
\def \As {{A\kern-.5em / \kern.5em}}
\def \Ds {D\kern-.7em / \kern.5em}

\def \g {\gamma}

\def \lam {\lambda}

\def \th {\theta}

\def \A {q}
\def \b {c_4}
\def \c {a_4}

\newcommand{\detail}[1]{}
\newcommand{\hide}[1]{}

\begin{document}

\pagestyle{plain}

%\begin{CJK}{UTF8}{bsmi} 

\begin{titlepage}
\vspace*{-10mm}   
\baselineskip 10pt   
\begin{flushright}   
\begin{tabular}{r} 
%KUNS-nnn\\
%RIKEN-iTHEMS-Report-18 
%April 1, 2018
\end{tabular}   
\end{flushright}   
\baselineskip 24pt   
\vglue 10mm

\begin{center}

\noindent
\textbf{\LARGE
Back Reaction of 4D Conformal Fields
\vskip10pt
on Static Black-Hole Geometry
}
\vskip20mm
\baselineskip 20pt

\renewcommand{\thefootnote}{\fnsymbol{footnote}}

{\large
Pei-Ming~Ho${}^a$
\footnote[1]{pmho@phys.ntu.edu.tw},
Hikaru~Kawai$^b$
\footnote[2]{hkawai@gauge.scphys.kyoto-u.ac.jp},
Yoshinori Matsuo${}^a$
\footnote[3]{matsuo@phys.ntu.edu.tw} and
Yuki~Yokokura$^c$
\footnote[4]{yuki.yokokura@riken.jp}
}

\renewcommand{\thefootnote}{\arabic{footnote}}

\vskip5mm

{\it  
${}^{a}$
Department of Physics and Center for Theoretical Physics, \\
National Taiwan University, Taipei 106, Taiwan,
R.O.C. 
\\
${}^{b}$
Department of Physics, Kyoto University, 
 Kitashirakawa, Kyoto 606-8502, Japan
 \\
${}^{c}$
iTHEMS Program, RIKEN, Wako, Saitama 351-0198, Japan
}

\vskip 25mm
\begin{abstract}

Static, spherically symmetric black hole solutions
to the semi-classical Einstein equation are studied,
including the effect of the quantum energy-momentum tensor
for conformal matters with 4D Weyl anomaly.
Through both perturbative and non-perturbative methods,
we show that the quantum effect can play a crucial role 
in shaping the near-horizon geometry,
and that the existence of the horizon requires fine-tuning.

\end{abstract}
\end{center}

\end{titlepage}

\tableofcontents

\vskip24pt

\hrule

\vskip24pt

\setcounter{page}{1}
\setcounter{footnote}{0}
\setcounter{section}{0}

%\chapter{}

\section{Introduction}

Since Hawking's proposal \cite{Hawking:1974sw} that 
a black hole can completely evaporate
through Hawking radiation,
physicists have realized that quantum effects,
despite its weakness,
have the potential to affect 
the large-scale structure of black holes.
However, 
to this day,
there has not yet been a satisfactory understanding on this topic,
leaving many unsettled issues,
including most notably the information loss paradox
\cite{Hawking-info-loss,Mathur:2009hf,Marolf:2017jkr}
and related proposals 
such as the fuzzball \cite{fuzzball} and the firewall \cite{firewall}.
Many believe that a rigorous detailed analysis
including the quantum effect is imperative.

Hence we aim to provide 
a rigorous detailed description of the black-hole geometry
with the back reaction of quantum fields 
taken into consideration.
In this paper,
we focus on static, spherically symmetric black holes.
We will give explicit expressions of
general solutions to the semi-classical Einstein equation.
Following Christensen and Fulling \cite{Christensen:1977jc},
we assume that the underlying quantum fields
responsible for the quantum energy-momentum tensor
are 4D conformal matters.
Unlike its 2D analogue \cite{Davies:1976ei},
its energy-momentum tensor is not uniquely fixed
by the trace anomaly and conservation law.
Instead of making additional assumptions to uniquely determine
the energy-momentum tensor,
we keep its full generality in our analysis.

%In the work of Davies, Fulling and Unruh \cite{Davies:1976ei},
%they analyzed the energy-momentum tensor of 
%a 2D massless scalar field in the Schwarzschild background.
%The result was later extended to 4D conformal matter fields
%by Christensen and Fulling \cite{Christensen:1977jc}.
%They found that there are different classes of vacuum states.
%The energy-momentum tensor can be either small or large 
%at the horizon depending on the vacuum states.
%The back reaction of the energy-momentum tensor
%on the black-hole geometry was however not analyzed.

%There are many misconceptions about black holes.

There are numerous related works in the literature.
Let us comment on some of those closely related.
Vacuum energy-momentum tensors derived from 
2D models of quantum field theories are 
extensively studied in Ref.\cite{Ho:2017joh}.
It was shown that,
depending on the quantum model of vacuum energy-momentum tensor
and the vacuum state,
the back-reacted near-horizon geometry
falls into three qualitatively different classes.
In one of the three classes,
the back reaction of quantum fields is insignificant,
while the event horizon is removed in the other two classes.

In one of the two classes that are horizonless,
the horizon is replaced by
a local minimum of the areal radius,
resembling the throat of a traversable wormhole \cite{Ho:2017joh}.
Similar results were also obtained in Refs.\cite{Fabbri}.
(The resemblance between the black holes and wormholes
was also noted in Refs.\cite{wormhole-as-BH}
based on different reasonings.)
The static geometry of the interior space
with a star composed of an incompressible fluid or a thin shell
was studied in Ref.\cite{Ho:2017vgi}.
The dynamical case including the effect of Hawking radiation
was explored via numerical simulation in Ref.\cite{Parentani:1994ij},
and then analytically in Ref.\cite{Ho:2018jkm}.

In the other horizonless class of solutions,
there is neither a horizon
nor a wormhole-like structure \cite{Ho:2017joh}.

The progress achieved in this work is mainly
the use of 4D (instead of 2D) models of quantum vacuum energy-momentum,
and its generality that covers all static solutions
with spherical symmetry.
All three classes of solutions found in various 2D models of vacuum energy 
are present in this 4D model,
and we note that the presence of the event horizon requires fine tuning.
The back reaction due to 4D conformal fields
has also been studied in Ref.\cite{Berthiere:2017tms},
but it was done in a manner different from this work,
and only the wormhole-like class was discussed.
Furthermore,
we emphasize the mathematical rigor of our results,
with our calculations
%are
carried out in both perturbative and non-perturbative approaches.

The plan of this paper is as follows.
We first lay out in Sec.\ref{Assumptions}
the assumptions behind the mathematical formulation we
%will
use to determine the black-hole geometry.
The general perturbative solution at the first order 
is given in Sec.\ref{generalT}, Sec.\ref{perturbation-z}
and Sec.\ref{near-Schwarzschild}
in three different coordinate systems,
each with its advantages and disadvantages.
The non-perturbative analysis is carried out in Sec.\ref{nonpert-analysis}.
The result is consistent with the perturbative solution.
The perturbative and non-perturbative solutions together
depict a comprehensive picture of the black-hole geometry.

\section{Semi-Classical Einstein Equation and 4D Conformal Matter}
\label{Assumptions}

\subsection{Semi-Classical Einstein Equation}

In this section,
we define the theoretical framework on which 
the analysis in this paper is based.
It is essentially Einstein's theory of gravity
sourced by 4D conformal quantum fields
through its expectation value of the quantum energy-momentum operator.

First,
we assume that the space-time geometry is determined 
by the semi-classical Einstein equation 
\be
G_{\mu\nu} = \kappa \langle T_{\mu\nu} \rangle
\label{scEE}
\ee
at large scales.
Here $\langle T_{\mu\nu} \rangle$ is 
the expectation value of the quantum energy-momentum operator $T_{\mu\nu}$
in the underlying quantum field theory.
A priori $\langle T_{\mu\nu} \rangle$ does not have 
to be the {\em vacuum} expectation value.
But in the perturbative calculation,
we will assume that $\langle T_{\mu\nu} \rangle$ is of $\mathcal{O}(\kappa^0)$,
so that the right hand side of eq.\eqref{scEE} vanishes 
in the classical limit $\kappa \rightarrow 0$.
In this sense, 
it is a vacuum expectation value which comes purely from quantum effects.
On the other hand,
in the non-perturbative analysis,
it can be the expectation value of an arbitrary state.

In this work,
we further assume that the energy-momentum tensor 
$\langle T_{\mu\nu} \rangle$ in eq.\eqref{scEE}
is given as that of 4D conformal quantum fields.
The advantage of considering conformal matters is that
$\langle T_{\mu\nu} \rangle$ is constrained by the Weyl anomaly,
leaving fewer uncertainties in $\langle T_{\mu\nu} \rangle$,
which is typically difficult to evaluate directly.

For 4D conformal quantum fields,
the trace of the energy-momentum tensor is given by the 4D Weyl anomaly
\be
\langle T^{\mu}{}_{\mu} \rangle = \b {\cal F} + \c {\cal G},
\label{anomaly}
\ee
which depends on two conformal charges $\b$ and $\c$
characterizing the conformal fields.
Here
\begin{align}
{\cal F} &\equiv
C^{\mu\nu\lambda\rho}C_{\mu\nu\lambda\rho}
=
R^{\mu\nu\lambda\rho}R_{\mu\nu\lambda\rho}
- 2 R^{\mu\nu}R_{\mu\nu} + \frac{1}{3} R^2,
\label{F}
\\
{\cal G} &\equiv
R^{\mu\nu\lambda\rho}R_{\mu\nu\lambda\rho}
- 4 R^{\mu\nu}R_{\mu\nu} + R^2,
\label{G}
\end{align}
where $C$ is the Weyl tensor
and ${\cal G}$ is the Gauss-Bonnet term.

The last assumption we shall make in this paper 
is that the configurations under study are 
static and spherically symmetric.
Locally, the metric can be put in the form
\be
ds^2 = - e^{\rho(r)} \left[dt^2 - \frac{dr^2}{F(r)}\right] + r^2 d\Omega^2.
\label{metric-1}
\ee
The coordinate $r$ is called the ``areal radius'',
in terms of which the area of a symmetric sphere is always $4\pi r^2$.
It is not necessarily monotonically increasing
in the radial direction.
We also define 
a ``proper radial coordinate'' $z$ by
\be
dz^2 = \frac{e^{\rho(r)}}{F(r)} dr^2.
\label{z-r}
\ee
The metric can then be expressed as
\be
ds^2 = - \frac{e^{A(z)}}{B^2(z)} dt^2 + dz^2 + B(z) d\Omega^2,
\label{metric-2}
\ee
where the two parametric functions $A(z)$ and $B(z)$
are related to $\rho(r)$ and $F(r)$ by eq.\eqref{z-r}
and
\be
r^2 = B(z),
\qquad
e^{\rho(r)} = \frac{e^{A(z)}}{B^2(z)}.
\ee

For a static, spherically symmetric configuration,
the only non-vanishing components of 
the energy-momentum tensor $\langle T_{\mu\nu} \rangle$ are
\be
\langle T^{t}{}_{t} \rangle, 
\quad
\langle T^{r}{}_{r} \rangle = \langle T^{z}{}_{z} \rangle,
\quad
\langle T^{\th}{}_{\th} \rangle = \langle T^{\phi}{}_{\phi} \rangle.
\ee
The three independent components of the energy-momentum tensor
(say, $\langle T^{t}{}_{t} \rangle, 
\langle T^{r}{}_{r} \rangle$
and $\langle T^{\th}{}_{\th} \rangle$)
are constrained by the conservation law 
\be
\nabla_{\mu}\langle T^{\mu}{}_{\nu}\rangle = 0,
\label{conservation}
\ee
as well as the anomaly equation \eqref{anomaly}.
There is thus only one independent functional degree of freedom
in the energy-momentum tensor.
We can arbitrarily specify $\langle T^r{}_r \rangle$
(or $\langle T^z{}_z \rangle$)
to be any given function of $r$ (or $z$)
from which all other components of $\langle T_{\mu\nu} \rangle$ 
are fixed.

The metric \eqref{metric-1} (or \eqref{metric-2})
has two independent functional parameters $\rho(r)$ and $F(r)$
(or $A(z)$ and $B(z)$)
to be solved from the semi-classical Einstein equation \eqref{scEE}.
Correspondingly,
only two of the 10 equations in eq.\eqref{scEE} are independent.
For conformal matters,
it is convenient to take the trace of the semi-classical Einstein equation
\begin{align}
G^{\mu}{}_{\mu} &= \kappa \langle T^{\mu}{}_{\mu} \rangle,
\label{Eq-trace}
\end{align}
where $\langle T^{\mu}{}_{\mu} \rangle$
is given by the Weyl anomaly \eqref{anomaly},
as one of the two independent equations.
In the following,
we will take the other independent equation to be
\begin{align}
G^{r}{}_{r} &= \kappa \langle T^{r}{}_{r} \rangle,
\label{Eq-rr}
\\
&\mbox{or}
\nn
\\
G^{z}{}_{z} &= \kappa \langle T^{z}{}_{z} \rangle,
\label{Eq-zz}
\end{align}
depending on our choice of coordinates.

\subsection{Energy-Momentum Tensor and Weyl Anomaly}\label{ssec:EM-anomaly}

%In the Schwarzschild background,
The most general static, spherically symmetric stress tensor
constrained by the conservation law
has only two functional degrees of freedom.
Following Christensen and Fulling \cite{Christensen:1977jc},
the energy-momentum tesnor can be parametrized by the trace
$\langle T^{\mu}{}_{\mu}\rangle$ and 
\be
\Theta \equiv \langle T^{\th}{}_{\th}\rangle
- \frac{1}{4}\langle T^{\mu}{}_{\mu}\rangle.
\label{Theta}
\ee
In addition, there are 2 dimensionless constant parameters $K$ and $Q$ which are a priori of order 1,
corresponding to different choices of boundary conditions. 

For the Schwarzschild background \eqref{metric-1} with 
\begin{align}
 \rho(r) &= \log\left(1-\frac{a}{r}\right), 
&
 F(r) &= \left(1-\frac{a}{r}\right)^2, 
\label{SS}
\end{align}
we obtain the following expression for 
the most general
time-independent, conserved and spherically symmetric stress tensor 
by integrating the conservation law; 
\begin{align}
\langle T^{t}{}_{t}\rangle &=
- \frac{1}{r(r-a)}
\left[\A + H(r) + G(r)\right]
+ \frac{1}{2} \langle T^{\mu}{}_{\mu}(r)\rangle - 2\Theta(r),
\label{Ttt}
\\
\langle T^{r}{}_{r}\rangle &=
\frac{1}{r(r-a)}
\left[\A + H(r) + G(r)\right],
\label{Trr}
\\
\langle T^{t}{}_{r}\rangle &=
\frac{1}{(r-a)^2}k,
\label{Ttr}
\\
\langle T^{\th}{}_{\th}\rangle &=
\Theta(r) + \frac{1}{4} \langle T^{\mu}{}_{\mu}\rangle,
\label{Tthth}
\end{align}
while
\begin{align}
\langle T^{r}{}_{t}\rangle &= 
- \left(1 - \frac{a}{r}\right)^2 \langle T^{t}{}_{r}\rangle,
\label{Trt}
\\
\langle T^{\phi}{}_{\phi}\rangle &=
\langle T^{\th}{}_{\th}\rangle, 
\label{Tphiphi}
\end{align}
and the two functions $H$ and $G$ are defined by
\begin{align}
H(r) &=
\frac{1}{2} \int_{a}^{r} \left(r'-\frac{a}{2}\right)
\langle T^{\mu}{}_{\mu}(r')\rangle dr',
\label{H-def}
\\
G(r) &= 2 \int_{a}^{r} \left(r'-\frac{3a}{2}\right)\Theta(r')dr'.
\label{G-def}
\end{align}
The integration constants $\A$ and $k$ are related 
to the parameters $Q$ and $K$ in Ref.\cite{Christensen:1977jc}
via 
\begin{align}
\A &= \frac{4(Q-K)}{a^2},
\label{A-def}
\\
k &= \frac{4K}{a^2},
\label{k-def}
\end{align}
which are of order $\mathcal{O}(a^{-2})$.

%YM-8/27
%The choice of the relative sign of $K$ in eqs.\eqref{A-def} and \eqref{k-def}
%is made such that the parameter $K$ does not lead to
%a divergent energy flux %$\langle T_{uu}\rangle (1-a/r)^{-2}$
%for a generic orthonormal frame at $r = a$.
%On the other hand,
Here, $Q=0$, or equivalently, $q=-k$ does not lead to 
the divergent outgoing energy flux but the 
incoming energy flux diverges at the past horizon, 
while the condition $q=k$ gives divergence of 
the outgoing energy flux at the future horizon 
but no divergence in incoming energy flux. 
The condition $q=k=0$, gives no divergent energy flux  
while $q=0$ generally provides non-zero energy flux in the asymptotic region. 
For example, the energy-momentum tensor which satisfies the conditions 
$\langle T^\theta{}_\theta\rangle =0$ and the anomaly condition \eqref{anomaly}
has the following asymptotic behavior; 
\begin{align}
\langle T^{t}{}_{t}(r \rightarrow \infty)\rangle &=
- \frac{a^2\A + \frac{6}{5}(\b+\c)}{a^2 r^2}
+ \mathcal{O}(r^{-3}),
\\
\langle T^{r}{}_{r}(r \rightarrow \infty)\rangle &=
 \frac{a^2\A + \frac{6}{5}(\b+\c)}{a^2 r^2}
+ \mathcal{O}(r^{-3}) \ . 
\end{align}
The condition $q=k=0$ corresponds to 
the Hartle-Hawking vacuum,
in which the incoming and outgoing energy fluxes 
are finite but balanced with each other. 
For $\A = - \frac{6(\b+\c)}{5 a^2}$ and $k=0$, 
the incoming and outgoing energy fluxes in the asymptotic region are zero 
but there is divergence at $r=a$. 
This condition corresponds to the Boulware vacuum. 
%This was the reason why 
%the Unruh vacuum and 
The Hartle-Hawking vacuum was considered physical,
and the Boulware vacuum unphysical,  
because of the divergence in the energy flux. 
%(The Boulware vacuum has $Q \neq 0$ 
%and the other two vacua have $Q = 0$.)
%-8/27
In this work,
we do not jump into the same conclusion before
closely examining the solutions.

Note that if $k \neq 0$,
the energy-momentum tensor breaks the time-reversal symmetry,
so the metric will not be static,
even though the energy-momentum tensor is time-independent.
We shall assume that $k = 0$
except briefly commentting on the case $k \neq 0$ in Sec.\ref{time-dependent}.

Explicit expressions of the energy-momentum tensor 
can now be given as follows.
For the Schwarzschild background,
\begin{align}
R^{\mu\nu\lambda\rho}R_{\mu\nu\lambda\rho} = \frac{12 a^2}{r^6},
\qquad
R^{\mu\nu}R_{\mu\nu} = 0,
\qquad
R = 0.
\end{align}
As a result,
\be
{\cal F} = {\cal G} = \frac{12 a^2}{r^6},
\ee
and the trace anomaly \eqref{anomaly} is
\be
\langle T^{\mu}{}_{\mu} \rangle = \frac{12(\b+\c)a^2}{r^6}.
\label{pert-trace}
\ee
At the first order in the perturbation theory,
this is the only place where the conformal charges 
appear in the semi-classical Einstein equation,
hence they only appear in the combination
$(\b+\c)$ in the first order perturbative solution.

For the Weyl anomaly \eqref{pert-trace},
we have
\begin{align}\label{HSch}
H(r) &=
\frac{3(\b+\c)(3r^5 - 5a^4 r + 2a^5)}{10 a^2 r^5},
\\
\label{GSch}
G(r) &=
\frac{3(\b+\c)(r^5 + 5a^4 r - 6a^5)}{10 a^2 r^5}
+ 2 \int_{a}^{r} \left(r'-\frac{3a}{2}\right)
\langle T^{\th}{}_{\th}(r')\rangle dr'.
\end{align}
One can thus compute the energy-momentum tensor 
using eqs.\eqref{Ttt}--\eqref{Tthth}
for the Schwarzschild background.

The expressions above given in terms of the areal radius $r$
can be easily generalized to 
an arbitrary radial coordinate $y$ for
the most general static, spherically symmetric background, 
\begin{equation}
 ds^2 = - e^{\rho(y)} dt^2 + h(y) dy^2 + r^2(y) d \Omega^2.
\end{equation}
The radial component of the metric $h(y)$ 
is related to those in other coordinates as 
\begin{equation}
 h(y) dy^2 = \frac{e^{\rho(r)}}{F(r)}dr^2 = dz^2. 
\end{equation}
For instance, $h(y) = 1 $ for $y=z$,
and $h(y) = \frac{e^{\rho(r)}}{F(r)}$ for $y=r$. 
By integrating the conservation equations, 
the energy-momentum tensor is found to be 
\begin{align}
\langle T^{t}{}_{t}\rangle &=
- \frac{e^{-\rho(y)}}{r^2(y)}
\left[\A + H(y) + G(y)\right]
+ \frac{1}{2} \langle T^{\mu}{}_{\mu}(y)\rangle - 2\Theta(y),
\\
\langle T^{y}{}_{y}\rangle &=
\frac{e^{-\rho(y)}}{r^2(y)}
\left[\A + H(y) + G(y)\right],
\label{Tyy}
\\
\langle T_{ty}\rangle &=
%%% pm %%%
-
%%%
\frac{k\,e^{-\rho(y)/2} \sqrt{h(y)}}{r^2(y)},
\\
\langle T^{\th}{}_{\th}\rangle &= \langle T^{\phi}{}_{\phi}\rangle = 
\Theta(y) + \frac{1}{4} \langle T^{\mu}{}_{\mu}\rangle,
\end{align}
where $H$ and $G$ are given by
\begin{align}
H(y) &= \frac{1}{4} G(y) + 
\frac{1}{2} \int_{\bar y}^{y} e^{\rho(y')} \rho'(y') r^2(y') 
\langle T^{\mu}{}_{\mu}(y')\rangle dy',
\\
G(y) &=
\int_{\bar y}^{y} e^{\rho(y')} r(y') \left(2 r'(y') - \rho'(y') r(y') \right)\Theta(y')dy'.
\end{align}
The primes on a variable (e.g. $\rho'$ and $r'$) indicate
the derivative with respect to the radial coordinate $y$. 
%The constant $a$ comes from the integration constant and related to $z_0$. 
%%%% pm %%%%
The constant $\bar y$ is the location of the horizon where $\rho\to-\infty$.
The divergence of the energy-momentum tensor at the horizon
is parametrized by the constant $q$ (with $H(y)+G(y)$ vanishing at $y = \bar{y}$), 
assuming that $\Theta(y)$ is finite there.
%Or, equivalently, $\rho\to-\infty$ in the limit $y\to \bar y$.%
\footnote{%
%Although $G(y)$ and $H(y)$ are expected to be $\mathcal O(y-\bar{y})$
%in the limit $y \rightarrow \bar{y}$.
In general, 
$r'(y)$ and $\rho'(y)$ could diverge as $y\to\bar y$ if $e^{\rho(y)}\to 0$ there. 
We assume that the integrands are still finite at $y = \bar y$, 
as in the case of the Schwarzschild metric. 
%%%% mp %%%%
}

For instance,
in terms of $A(z)$, $B(z)$ and the proper radial coordinate $z$
(see eq.\eqref{metric-2}), 
$H$ and $G$ are
\begin{align}
H(z) &= 
\frac{1}{4} \int_{\bar z}^{z} \frac{e^{A(z')}}{B(z')}
\left(A'(z') - \frac{B'(z')}{B(z')}\right)
\langle T^{\mu}{}_{\mu}(z')\rangle dz',
\label{H-def-nonpert}
\\
G(z) &= 
- \int_{\bar z}^{z} \frac{e^{A(z')}}{B(z')}
\left(A'(z') - 3\frac{B'(z')}{B(z')}\right)
\Theta(z')dz'.
\label{G-def-nonpert}
\end{align}

The metric for a static, spherically symmetric space-time
can in principle be solved from 
the semi-classical Einstein equation \eqref{scEE}
for an arbitrary assignment of $\langle T^\theta{}_\theta \rangle$.
In the following, 
we shall find the solution of the metric both perturbatively 
and non-perturbatively for arbitrary $\langle T^\theta{}_\theta \rangle$.
In the perturbative analysis,
we find the first order correction to the Schwarzschild metric.
The non-perturbative analysis is carried out
in a small neighborhood of an arbitrary point in space.
Putting the perturbative and non-perturbative results together,
we get a consistent picture of the black-hole geometry
including the back reaction of the energy-momentum tensor
of 4D conformal quantum fields.

\section{General Perturbative Solution in Areal Radius}
\label{generalT}

In the following,
we will solve the first order perturbative correction to 
the Schwarzschild metric due to the energy-momentum tensor
of 4D conformal quantum fields with full generality,
and we will classify the solutions according to 
their near-horizon geometry.
In this section,
we shall use the areal radius $r$
as the coordinate parametrizing the radial direction.
In the next two sections we will use other radial coordinates.

\subsection{Perturbative Analysis}
\label{general-EMT-pert}

Here we consider the ansatz \eqref{metric-1} for the metric.
The parametric functions $\rho(r)$ and $F(r)$
are expanded in powers of the Newton constant $\kappa$ as
\begin{align}
\rho &=
\rho_0 + \kappa \rho_1 + \kappa^2 \rho_2 + \cdots,
\label{rho-expansion}
\\
F &=
F_0 + \kappa F_1 + \kappa^2 F_2 + \cdots,
\label{F-expansion}
\end{align}
where $\rho_n$ and $F_n$ are of $\mathcal{O}(\kappa^0)$. 
Note that
$[\kappa]=L^2$ and $[\rho_1]=[F_1]=L^{-2}$
in terms of the dimension of length $L$.
The 0-th order solution is given by
the Schwarzschild metric:
\begin{align}
\rho_0 &= \log\left(1 - \frac{a}{r}\right),
\label{rho0}
\\
F_0 &= \left(1 - \frac{a}{r}\right)^2,
\label{F0}
\end{align}
where $a$ is the Schwarzschild radius.
We shall solve the semi-classical Einstein equation \eqref{scEE}
perturbatively in the $\kappa$-expansion 
for the leading order perturbative correction $\rho_1$, $F_1$.
%YM-6/27
We assume that the energy-momentum tensor comes from quantum effects 
and are of $\mathcal O(\kappa^0)$. 
%-6/27

For any given radial pressure $\langle T^r{}_r \rangle$,
the metric correction $\rho_1$ and $F_1$ can then be directly solved from
the semi-classical Einstein equations \eqref{Eq-rr} and \eqref{Eq-trace}.
Eq.\eqref{Eq-rr} is
\be
\frac{F_1(r)}{(r-a)^2} + \frac{(r-a) \rho'_1 - \rho_1}{r^2} 
= \langle T^r{}_r \rangle.
\ee
It allows one to express $F_1$ in terms of $\rho_1$:
\be
F_1(r) = (r-a)^2\left[
\langle T^r{}_r \rangle - \frac{(r-a) \rho'_1 - \rho_1}{r^2}
\right].
\label{F1-from-rho1}
\ee
Plugging this into the other equation \eqref{Eq-trace},
we turn it into a second order differential equation for $\rho_1$ only:
\begin{align}
(2r-3a)(r-a)\rho''_1 + 2(2r-3a)\rho'_1 
%- r^2(4r-3a) \langle (T^r{}_r)' \rangle
%- 6r(2r-a) \langle T^r{}_r \rangle
%+ 2r^2 \langle T^{\mu}{}_{\mu} \rangle
= J(r),
\end{align}
where
\be
J(r) \equiv
r^2(4r-3a) \langle (T^r{}_r)' \rangle
+ 6r(2r-a) \langle T^r{}_r \rangle
- 2r^2 \langle T^{\mu}{}_{\mu} \rangle.
\ee
We can then solve $\rho_1$ from this second order differential equation as
\begin{align}\label{rho1_r}
\rho_1(r) &=
C_0 + \int_{r_0}^r dr' \; \frac{1}{(r'-a)^2} \left[
C_1 + \int_{r_0}^{r'} dr'' \; \frac{(r''-a)^2 J(r'')}{(2r''-3a)(r''-a)}
\right],
\end{align}
with integration constants $C_0$ and $C_1$.
After finding the solution of $\rho_1$,
one can easily compute $F_1$ from eq.\eqref{F1-from-rho1}.

The first order corrections in the $\kappa$-expansion, 
$\rho_1(r)$ and $F_1(r)$,
are now written in terms of 
$\langle T^r{}_r\rangle$ and $\langle T^\mu{}_\mu \rangle$. 
The trace of the energy-momentum tensor $\langle T^\mu{}_\mu \rangle$ 
is given by \eqref{pert-trace} for the Schwarzschild background, 
while $\langle T^r{}_r\rangle$ is expressed as \eqref{Trr}
(with \eqref{HSch} and \eqref{GSch})
which diverges in general at $r=a$. 
Since this divergence is related to the coordinate singularity in the $(t,r)$-coordinates, 
%though it cannot be removed by using non-singular coordinates, 
we assume that $\langle T^\theta{}_\theta\rangle$,
which is invariant under the coordinate transformation in the $(t,r)$-directions,
does not diverge at $r=a$. 
Under this assumption, it would be convenient to express the results 
in terms of $\langle T^\theta{}_\theta\rangle$, or equivalently, of $\Theta$. 

%%% pm %%%
%In the limit $r \rightarrow a$,
%the energy-momentum tensor is given by
To calculate the energy-momentum tensor in the limit $r \rightarrow a$,
we expand eqs.\eqref{Ttt}, \eqref{Trr}, \eqref{HSch} and \eqref{GSch}
in powers of $(r-a)$ and find
%%%
%\begin{align}
%H(r\rightarrow a) &=
%\frac{3(\b+\c)}{a^3} (r-a) + \mathcal{O}((r-a)^2),
%\\
%G(r\rightarrow a) &=
%\left[\frac{3(\b+\c)}{a^3} - a \langle T^{\th}{}_{\th}\rangle\right] (r-a)
%+ \mathcal{O}((r-a)^2),
%\end{align}
%so that
\begin{align}
\langle T^{t}{}_{t}(r\rightarrow a)\rangle  &=
- \frac{\A}{a (r-a)}
%%% pm %%%
%- \frac{6(\b+\c)}{a^4}
%+ \langle T^{\mu}{}_{\mu}(r\rightarrow a)\rangle 
+ \frac{\A + 6(\b+\c)}{a^4}
%%%
- \langle T^{\th}{}_{\th}(r\rightarrow a)\rangle ,
\label{Ttt-r=a}
\\
\langle T^{r}{}_{r}(r\rightarrow a)\rangle  &=
\frac{\A}{a(r-a)}
%%% pm %%%
%+ \frac{6(\b+\c)}{a^4} 
+ \frac{-\A + 6(\b+\c)}{a^4} 
%%%
- \langle T^{\th}{}_{\th}(r\rightarrow a)\rangle .
\label{Trr-r=a}
%\\
%T^{t}{}_{r}(r\rightarrow a) &=
%\frac{1}{(r-a)^2}\frac{4K}{a^2},
%\\
%T^{\th}{}_{\th}(r\rightarrow a) &=
%\Theta(r) + \frac{1}{4} T^{\mu}{}_{\mu},
\end{align}
%Notice that $\langle T^r{}_r(r) \rangle$ is composed of a diverging term in the limit.
We shall thus represent $\langle T^r{}_r \rangle$ 
(for all $r$, not only for $r$ close to $a$) as
\be
\langle T^r{}_r(r) \rangle = \frac{q}{a(r-a)} + f(r),
\label{Tupdownrr}
\ee
where
\be
f(r) \equiv
- \frac{\A}{ar}
+ \frac{1}{r(r-a)}
\left[H(r) + G(r)\right].
\label{f-pert}
\ee
Here, $H(r)$ and $G(r)$ are given by \eqref{HSch} and \eqref{GSch}, respectively.
The function $f(r)$ is determined by $\langle T^\theta{}_\theta\rangle$ and $q$,
and it is regular at $r = a$
(assuming that $\langle T_{\th\th}(a) \rangle$ is finite). 
%YM-6/27
As the energy-momentum tensor is assumed to come from quantum effects,
$q$ should be of $\mathcal O(\kappa^0 a^{-2})$. 
The results \eqref{F1-from-rho1} \eqref{rho1_r} are now expressed in terms of $f(r)$ as 
\begin{align}
\rho_1(r) &=
C_0 + \frac{C_1}{r-a} 
+ W(r)
\nn \\
&
- \frac{\A}{4a(r-a)} \left[
2(7a^2 + 8ar - 4r^2)
+ a(3a-2r) \left(
4\log\left(\frac{r-a}{a}\right) + 9\log\left(\frac{3a-2r}{a}\right)
\right)
\right],
\label{rho1-generalT}
\\
F_1(r) &=
- (r-a) \Big\{
\frac{\left[
- 2 C_0 (r - a) - 4 C_1
\right]}{2 r^2}
\nn \\
&
+ \frac{
\A \left[
7a^2 - 2r^2 + 13ar
+ 9a(2a - r)\log\left(\frac{2r-3a}{a}\right)
+ 4a(2a - r)\log\left(\frac{r-a}{a}\right)
\right]
}{2a r^2}
\nn \\
& 
- (r-a) f(r)
+ \frac{1}{2r^2}\Big[
V(r) - 2(r-a)W(r)
\Big]
\Big\},
\label{F1-generalT}
\end{align}
where
\begin{align}
W(r) &\equiv
\int_a^r dr'' 
\frac{1}{(r''-a)^2} \frac{V(r'')}{2},
%\int_a^{r''} dr' 
%\frac{\kappa r'(r'-a)}{2r' - 3a}
%\left[
%6(2r'-a)f(r') + r'(4r'-3a)f'(x) - 2r' T^{\mu}{}_{\mu}
%\right]
\hide{%%% Beginning of \hide{}
\nn \\
&=
- \frac{1}{2} a(r-a) \left[
6 f(a) + a f'(a) - 2 \langle T^{\mu}{}_{\mu}(a)\rangle
\right] + \cdots.
}%%% end of \hide{}
\\
V(r) &\equiv
- 2\int_a^r dr' \frac{r'(a-r')}{(3a-2r')}
\left[ 6(a-2r')f(r') + 2r' \langle T^{\mu}{}_{\mu}(r')\rangle 
+ r'(3a-4r')f'(r')\right].
\hide{%%% Beginning of \hide{}
\nn \\
&=
\kappa \left[
\frac{(\b+\c)}{a^3} - 6a f(a) - a^2 f'(a)
\right] (r-a)^2 + \mathcal{O}((r-a)^3).
}%%% end of \hide{}
\end{align}
%Note that the trace $\langle T^{\mu}{}_{\mu}(a)\rangle$ and 
%the function $f(a)$ at $r = a$ are given by
%\begin{align}
%\langle T^{\mu}{}_{\mu}(a)\rangle &= \frac{12(\b+\c)}{a^4},
%\\
%f(a) &= - \frac{1}{a^4}\left[
%a^2\A + 9(\b+\c) \right].
%\end{align}

There are two integration constants $C_0$ and $C_1$.
The constant $C_0$ corresponds to a scaling of the time coordinate $t$;
and $C_1$ to a shift of the Schwarzschild radius $a$.
They are fixed by specifying the asymptotic conditions at $r\rightarrow\infty$:
the choice of $t$ in the asymptotically flat spacetime fixes $C_0$;
and the asymptotic total energy 
(which depends on the choice of $f(r)$)
determines $C_1$.

Eqs.\eqref{rho1-generalT} and \eqref{F1-generalT} give
the most general first order perturbation of the Schwarzschild metric
for a 4D conformal matter field in any state according to 
the semi-classical Einstein equation \eqref{scEE}.
The energy-momentum tensor $\langle T_{\mu\nu}\rangle$
of any quantum state is specified by
a function $f(r)$ and a constant $q$
through eq.\eqref{Tupdownrr},
with the rest of the energy-momentum tensor determined 
through the Weyl anomaly and conservation law.

\hide{%%% Beginning of \hide{}
At large $r$,
\begin{align}
V(r) &\simeq
4\langle T_{\th\th}\rangle r^3
+ \left(-\frac{4\A}{a} + 6a\langle T_{\th\th}(\infty)\rangle
+ 6\gamma_1\right)r^2
\nn \\
&
+ \left(\frac{24(\b+\c)}{5a^2}-\A
+ 18 a^2 T_{\th\th}(\infty) + 12\gamma_0 + 12 a\gamma_1\right) r
\nn \\
&
+ \left(\frac{12(\b+\c)}{5a}-\frac{9a\A}{2}
+ 27 a^3 T_{\th\th}(\infty) + 12 a\gamma_0 + 18 a^2\gamma_1\right) \log(r)
%+ c_v
\nn \\
&
- \left(\frac{12(\b+\c)}{5} - aG_1 - \frac{27}{4}a^2\A
+ \frac{81}{2}a^4 T_{\th\th}(\infty) + 18a^2\gamma_0 + 27a^3\gamma_1
\right) \frac{1}{r}
+ \mathcal{O}(1/r^2),
\\
W(r) &\simeq
T_{\th\th}(\infty)r^2
+ \kappa\left(-\frac{2\kappa\A}{a} + 7 aT_{\th\th}(\infty)
+ 3\gamma_1\right)r
\nn \\
&
+ \frac{3\kappa}{10a^2}\left(8(\b+\c) - 15a^2\A + 70a^4 T_{\th\th}(\infty)
+ 20a^2\gamma_0 + 40a^3\gamma_1\right) \log(r)
\nn \\
&
%+ c_w
- \frac{\kappa}{20a}\left(24(\b+\c)-45a^2\A
+270a^4 T_{\th\th}(\infty)+120a^2\gamma_0+180a^3\gamma_1\right) \frac{\log(r)}{r}
\nn \\
&
- \frac{\kappa}{20a}(%10ac_v+
96(\b+\c)-140a^2\A
+540a^4 T_{\th\th}(\infty)+240a^2\gamma_0+420a^3\gamma_1) \frac{1}{r}
+ \mathcal{O}(1/r^2),
\end{align}
}%%% end of \hide{}

%%% pm %%%
%\subsubsection{Analysis in $r\to\infty$}
\subsubsection{Analysis in the Limit $r\to\infty$}
%%%

Now we consider the solution \eqref{rho1-generalT}--\eqref{F1-generalT}
in the two limits $r \rightarrow \infty$ and $r \rightarrow a$.
In the asymptotic region at large $r$,
the asymptotic expressions of the metric is given by
\begin{align}
\rho_1(r) &\simeq
\langle T^{\theta}{}_{\th}(\infty)\rangle r^2
+ (7a\langle T^{\theta}{}_{\th}(\infty)\rangle+3\gamma_1)r
+ S \log(r)
\nn \\
&
+ \Big[C_0 %+ c_w 
- 2\A + \frac{9}{2}\A\log(2)
- \frac{13}{2}\A\log(a)\Big]
\nn \\
&
- \Big[
\frac{6(\b+\c)}{5a}+a\A
+ \frac{3a}{2}(9a^2 \langle T^{\theta}{}_{\th}(\infty)\rangle + 4\g_0 + 6a\g_1)
\Big]\frac{\log(r)}{r}
\nn \\
&
+ \frac{1}{a} \Big[
C_1
%- \frac{c_v}{2}
- \frac{24}{5a}(\b+\c)
- \frac{29}{4} a\A - 27a^3 \langle T^{\theta}{}_{\th}(\infty)\rangle
- 12a\g_0 - 21a^2\g_1
\nn \\
&
- \frac{9}{4}a\A\log(2) + \frac{13}{4}a\A\log(a)
\Big]\frac{1}{r}
+ \mathcal{O}\left(\frac{1}{r^2}\right),
\displaybreak[1]\\
%%% F1 remains to be checked against Mathematica. %%%
F_1(r) &\simeq
(3a\langle T^{\theta}{}_{\th}(\infty)\rangle + \g_1)r
+ S \log(r)
\nn \\
&
+ \Big[C_0 %+ c_w 
- \frac{6(\b+\c)}{5a^2} - 3\A
+ \frac{9}{2}\A\log(2) - \frac{13}{2}\A\log(a)
- 19a^4 \langle T^{\theta}{}_{\th}(\infty)\rangle 
- 5\g_0 - 10a\g_1\Big]
\nn \\
&
- \Big[
\frac{36(\b+\c)}{5a}+6a\A+69a^3 \langle T^{\theta}{}_{\th}(\infty)\rangle
+ 24a\g_0 + 42a^2\g_1\Big]\frac{\log(r)}{r}
\nn \\
&
+ \Big[
-2aC_0 + 2C_1 %- c_v - 2ac_w
- \frac{18(\b+\c)}{5a} - \frac{9a\A}{4} + \kappa G_1
\nn \\
&
- 11a^3 \langle T^{\theta}{}_{\th}(\infty)\rangle
- 7a\g_0 - 12a^2\g_1
- \frac{27a\A}{2}\log(2)
+ \frac{39a\A}{2}\log(a)
\Big]\frac{1}{r}
+ \mathcal{O}\left(\frac{1}{r^2}\right),
\end{align}
where
\be
S \equiv \frac{12(\b+\c)}{5a^2}
+ 2\A + 21a^2 \langle T^{\theta}{}_{\th}(\infty)\rangle + 6\gamma_0 + 12a\gamma_1.
\ee
The constant parameters $\g_0$, $\g_1$ and $G_1$
are defined by the large-$r$ expansion of $G(r)$ \eqref{GSch} as
\be
G(r) \simeq \langle T^{\theta}{}_{\th}(\infty)\rangle r^2
+ \gamma_1 r + \frac{3(\b+\c)}{10a^2} +
\gamma_0 + \frac{G_1}{r} + \mathcal{O}(1/r^2), 
\ee
and $\langle T^{\theta}{}_{\th}(\infty)\rangle$ is 
the angular pressure in asymptotically flat region, 
$\langle T^{\theta}{}_{\th}(r\to\infty)\rangle$. 
As the anomaly \eqref{pert-trace} goes to $0$ in the asymptotically flat region
\be
\langle T^{\mu}{}_{\mu}(r \rightarrow \infty)\rangle = 0,
\ee
the part independent of $r$ in $\langle T_{\mu\nu}(r\rightarrow\infty) \rangle$ 
represents a thermal equilibrium state at spatial infinity
parametrized by $\langle T^{\th}{}_{\th}(\infty)\rangle$.
%As we mentioned above,
%$\langle T^{\theta}{}_{\th}(\infty)\rangle$ parametrizes 
%the temperature of thermal equilibrium 
%at the spatial infinity.

%%% pm %%%
\subsubsection{Analysis in the Limit $r\to a$}
%%%

In the near-horizon limit $r \rightarrow a$,
we parametrize $C_0$ and $C_1$
in terms of two parameters $\tilde{c}_0$ and $\tilde{c}_1$
%$L = a e^{\tilde{c}_1/\kappa aq}$ 
defined by
\begin{align}
C_0 = \tilde{c}_0 - \frac{9}{2}\A,
\qquad
C_1 = \frac{9}{2}a\A 
+ \tilde{c}_1.
\end{align}
Here, $[C_0]=L^{-2}$ and $[C_1]=L^{-1}$ for the dimension of length $L$, 
and $\tilde c_0\sim\mathcal O(a^{-2})$ and $\tilde c_1\sim\mathcal O(a^{-1})$.
%Here $L$ can be determined by $C_v$.
Then the solution is approximated by
\begin{align}
\rho_1(r) &=
%C_0 + \frac{9}{2}\kappa\A(1+\log(a)) + 2\kappa\A\log(L)
%+ \frac{C_1 - \frac{11}{2}\kappa a\A - \frac{9}{4}\kappa a\A\log(a) 
%- \kappa a\A\log(L)}{r-a}
- \frac{\A a}{(r-a)}\log\left(\frac{r-a}{a}\right)
+ \frac{\tilde{c}_1 - \A a}{(r-a)}
+ 2\A \log\left(\frac{r-a}{a}\right) 
+ \tilde{c}_0
\nn \\
&
+ \left[- \frac{5}{2}\frac{\A}{a} - 3af(a) 
+ a\langle T^{\mu}{}_{\mu}(a)\rangle
- \frac{1}{2}a^2 f'(a) \right](r-a)
+ \mathcal{O}((r-a)^2),
\label{rho1}
\\
F_1(r) &= 
%\left[
%\frac{2C_1}{a^2} - \frac{11}{a}\kappa\A - \frac{9}{2a}\kappa\A\log(a)
%- \frac{2}{a}\kappa\A\log(L)
%\right](r-a)
\hide{%%% Beginning of \hide{}
- \frac{2\A (r-a)}{a} 
\log\left(\frac{r-a}{a}\right)
+ \frac{2\A (r-a)}{a}\frac{\tilde{c}_1}{\kappa aq}
+ \frac{2\A (r-a)}{a} 
\nn \\
&
+ \left[
\frac{C_0}{a^2} - \frac{4C_1}{a^3} + \frac{45}{2a^2}\A
+ f(a) + \frac{27}{2a^2}\A\log(a)
+ \frac{6}{a^2}\A\log(r-a)
\right](r-a)^2
+ \mathcal{O}((r-a)^3)
\nn \\
&=
}%%% end of \hide{}
- \frac{2\A (r-a)}{a}
\log\left(\frac{r-a}{a}\right)
+ \frac{2\tilde{c}_1}{a^2}(r-a)
\nn \\
&
+ \left[
\frac{6}{a^2}\A\log\left(\frac{r-a}{a}\right)
+ \frac{\tilde{c}_0}{r^2}
- \frac{4\tilde{c}_1}{a^3}
+ f(a)
\right](r-a)^2
+ \mathcal{O}((r-a)^3).
\label{F1}
\end{align}
From these expressions 
we see that,
in the near-horizon region,
the perturbation at the leading order depends
only on the Schwarzschild radius $a$
and the constant parameter $\A$,
but not on the conformal charges 
or the tangential pressure $\langle T_{\th\th}(a) \rangle$.

Clearly,
the perturbation theory fails at $r = a$
where $\rho_1$ diverges,
while it provides a good approximation at large $r$.
To find out the range of $r$ where 
the perturbation theory works,
we examine the order of magnitude of 
some geometric quantities at $r = a$:
\begin{align}
R &= - \kappa\left[
\frac{5q}{a^2}\log\left(\frac{r-a}{a}\right)
+ \mathcal{O}((r-a)^0)
\right]
+ \mathcal{O}(\kappa^2),
\\
R_{\mu\nu}R^{\mu\nu} &= \mathcal{O}(\kappa^2),
\\
R_{\mu\nu\lam\sigma}R^{\mu\nu\lam\sigma} &=
\frac{12a^2}{r^6}
+ \kappa\left[
\frac{2q}{a^4}\log\left(\frac{r-a}{a}\right)
+ \mathcal{O}((r-a)^0)
\right]
+ \mathcal{O}(\kappa^2).
\end{align}
%YM-6/27
Noting $q=\mathcal O(a^{-2})$,
the quantum correction of $R_{\mu\nu\lam\sigma}R^{\mu\nu\lam\sigma}$ 
is sufficiently small as long as 
\begin{equation}
 r - a \gg  a e^{-a^2/\kappa}. 
\end{equation}
Therefore, the perturbative expansion for the geometry is expected to be valid in this region. 
On the other hand, the perturbative corrections for $\rho$ and $F$, 
as is shown in \eqref{rho1} and \eqref{F1}, 
become comparable to the leading order terms around the region 
\be
r - a \sim \mathcal{O}\left(\frac{\kappa}{a}\right). 
\label{r-a-order}
\ee
This implies that the expressions \eqref{rho1} and \eqref{F1} 
are valid only for 
\be
r - a \gtrsim \mathcal{O}\left(\frac{\kappa}{a}\right),
\label{r-range}
\ee
since they are calculated by assuming that $r-a\sim \mathcal O(\kappa^0 a)$. 
The quantum corrections for geometric quantities above are all of
$\mathcal{O}(\kappa\log\kappa)$ or smaller in \eqref{r-range} %as long as
%\be
%r - a \gtrsim \mathcal{O}\left(\frac{\kappa}{a}\right),
%\label{r-range}
%\ee
so we expect that the perturbation theory is valid
in the range defined by eq.\eqref{r-range}.

Within the range of validity of the perturbation theory,
the quantum correction is most significant around the region \eqref{r-a-order} %where
%\be
%r - a \sim \mathcal{O}\left(\frac{\kappa}{a}\right),
%\label{r-a-order}
%\ee
%
%-6/27
although it can in principle be even larger in the non-perturbative domain.
The energy-momentum tensor \eqref{Ttt-r=a} and \eqref{Trr-r=a}
around the region \eqref{r-a-order} is estimated to be
\begin{align}\label{T_aa}
|\langle T^t{}_t \rangle|
%&\simeq - \frac{1}{\kappa a^2\zeta}
%- \langle T^{\th}{}_{\th}(\kappa qa\zeta) \rangle
%+ \mathcal{O}(\kappa^0)
%&\sim \mathcal{O}\left(\frac{1}{\kappa a^2}\right),
%\\
\sim
|\langle T^r{}_r \rangle|
%&\simeq \frac{1}{\kappa a^2\zeta}
%- \langle T^{\th}{}_{\th}(\kappa qa\zeta) \rangle
%+ \mathcal{O}(\kappa^0)
%&
\sim \mathcal{O}\left(\frac{1}{\kappa a^2}\right).
\end{align}
This is what one would expect,
according to the Einstein equation,
of the energy-momentum tensor for a spacetime region
in which the curvature is of $\mathcal{O}(1/a^2)$
\footnote{In a self-consistetn model
\cite{Kawai:2013mda,Kawai:2014afa,Ho:2015fja,Kawai:2015uya,Ho:2015vga,Ho:2016acf,Kawai:2017txu}, 
\eqref{T_aa} is obtained.}.
On the other hand, 
in comparison with the classical vacuum energy
(which is exactly zero),
the quantum vacuum energy is relatively high.
For a black hole of a few solar mass,
the energy density can be as high as $\mathcal{O}(\mbox{MeV}^4)$.

\subsection{Classification of Solutions}

To understand the quantum-corrected near-horizon geometry in more details,
we examine the perturbative solution in the neighborhood \eqref{r-range}.
According to \eqref{rho0} and \eqref{rho1},
we find
\begin{align}
\rho &= \rho_0 + \kappa\rho_1 + \cdots
%\nn \\&
= \left[1 - \frac{\kappa a\A}{(r-a)}\right]
\log\left(1 - \frac{a}{r}\right)
%+ \frac{\tilde{c}_1-\kappa \A a}{(r-a)}
%+ 2\kappa \A \log\left(\frac{r-a}{a}\right) 
+ \cdots,
\label{rho0+rho1}
\end{align}
and using \eqref{F0} and \eqref{F1},
we find
\begin{align}\label{F0+F1}
F &= F_0 + \kappa F_1 + \cdots
%\nn \\&
= \frac{(r-a)^2}{a^2} 
\left[1
- \frac{2\kappa a\A}{(r-a)} \log\left(\frac{r-a}{a}\right)
\right]
+ \cdots.
%+ \cdots
%\nn \\
%&\simeq
%\frac{r-a}{a^2} \left[r-a
%- 2\kappa a\A \log\left(\frac{r-a}{L}\right)
%\right].
\end{align}
These expressions determine the near-horizon geometry
to the first order in the perturbation theory.
The nature of the near-horizon geometry depends on
the sign of the parameter $\A$. 
%YM-8/27
The Hartle-Hawking vacuum corresponds to the condition $\A=0$, 
while $q$ would be negative in the Boulware vacuum, 
as we have seen in an example in Sec.~\ref{ssec:EM-anomaly}.
%-8/27
We consider the three possibilities:
$q < 0$, $q = 0$ and $q > 0$, separately in the following.

\subsubsection{Wormhole-Like Throat ($q < 0$)}
\label{throat-r}

If $q < 0$,
in the limit $r \rightarrow a$
%%% pm %%%
%from the side $r > a$,
(moving towards $r = a$ from distance),
%%%
we have
\begin{align}
\rho &\simeq \rho_0 + \kappa \rho_1 
\simeq \left[1 + \frac{\kappa|q|a}{r-a}\right]\log\left(\frac{r-a}{a}\right)
+ \cdots,
\\
F &\simeq F_0 + \kappa F_1
\simeq \frac{1}{a^2}(r-a)^2
\left[1 + \frac{2\kappa|q|a}{r-a}\log\left(\frac{r-a}{a}\right)\right] + \cdots.
\end{align}
The perturbation theory breaks down when
$r$ is too close to the Schwarzschild radius:
\be
r - a \ll \kappa|q|a.
\ee
Hence the existence of the horizon at $r = a$
is not guaranteed.

According to this approximation,
$F$ has a zero at $r > a$,
implying that there is a local minimum in $r$
outside the horizon.
\footnote{
To claim that the neck is a local minimum of $r$,
we 
%%% pm %%%
%need to check 
have checked
%%%
not only that $F = 0$ at the neck,
but also that 
%\be
$\left.\frac{dF}{dr}\right|_{\text{neck}} > 0$.
%\ee
%%% pm %%%
%Indeed, we can check the both.
%%%
%\be
%\left.\frac{dF}{dr}\right|_{\mbox{neck}}
%\simeq \frac{1}{a^2}(r-a),
%\ee
%which is positive since $r > a$ at the neck.
}
We refer to this local minimum in $r$ as the ``neck'' or ``throat''
of a ``wormhole-like structure'',
as it resembles the geometry of a traversable wormhole,
although it does not lead to another open spacetime.
Noting eq.\eqref{z-r}, the location of the wormhole neck is where
$dr/dz = 0$,
i.e. where
\be
F(r) = 0.
\ee
With $F(r)$ approximated by $F_0 + \kappa F_1$,
$r - a$ is of $\mathcal{O}(\kappa/a)$ at the neck.
But this means that the 0-th order, 1st order and 2nd order terms in $F$
are all of $\mathcal{O}(\kappa^2/a^4)$.
Hence it is not reliable to determine the location
of the neck based the first order perturbation in the $r$-coordinate.

In Sec.\ref{perturbation-z}, 
we will see that the perturbation theory in another coordinate system
(using the proper radial coordinate $z$)
allows us to locate the neck with better accuracy.
In Sec.\ref{nonpert-wormhole},
we will further confirm this perturbative result 
by non-perturbative analysis.

\subsubsection{Event Horizon ($q = 0$)}
\label{horizon-r}

When $q = 0$,
we have from eqs.\eqref{rho1-generalT} and \eqref{F1-generalT}
\begin{align}
\rho_1 &\simeq
- \frac{1}{(r-a)}\Delta a
%\Big[c_1 + \frac{3(\b+\c)}{a} + \frac{6(\b+\c)}{5a}\log(a)\Big] 
+ \cdots,
\\
F_1 &\simeq
- \frac{2}{a^2}
%\Big[c_1 + \frac{3(\b+\c)}{a} + \frac{6(\b+\c)}{5a}\log(a)\Big]
(r-a)\Delta a
+ \cdots,
\end{align}
where
\be
\Delta a \equiv
C_1 + 3(\b+\c)\frac{(5+2\log(a))}{5a}.
\ee
These expressions coincide with the perturbation of the Schwarzschild metric
for a shift of the Schwarzschild radius by
\be
a \rightarrow a + \Delta a.
\ee
This means that the black-hole mass receives
a quantum correction $\Delta a/2$.
We can redefine the Schwarzschild radius $a$
(tune $C_1$ such that $\Delta a = 0$)
to absorb this correction.
As a result,
$\rho_1$ and $F_1$ are only modified
at higher orders of the $(r-a)$-expansion
in the near-horizon region,
and the geometry of the Schwarzschild horizon
is not significantly modified by the quantum effect.
Only in this case,
the perturbation theory is valid down to the horizon at $r = a$.

%%% pm %%%
%\subsubsection{No Neck And No Horizon ($q > 0$)}
\subsubsection{No Neck \& No Horizon ($q > 0$)}
\label{nono-r}
%%%

If $q > 0$,
in the limit $r \rightarrow a$
%%% pm %%%
%from the side $r > a$,
(moving towards $r = a$ from distance),
we have
%%%
\begin{align}
\rho &\simeq \rho_0 + \kappa \rho_1 
\simeq \left[1 - \frac{\kappa|q|a}{r-a}\right]\log(r-a)
+ \cdots,
\\
F &\simeq F_0 + \kappa F_1
\simeq \frac{1}{a^2}(r-a)^2
\left[1 - \frac{2\kappa|q|a}{r-a}\log(r-a)\right] + \cdots,
\end{align}
from eqs.\eqref{rho0+rho1} and \eqref{F0+F1}.
In contrast with the case $q < 0$,
there is no local minimum of $r$ outside 
the Schwarzschild radius,
because $F$ has no zero for $r > a$,
while the perturbation theory breaks down in the region
\be
r - a \ll \kappa|q|a.
\ee

Furthermore,
since $\rho_1 \rightarrow + \infty$ in the limit $r \rightarrow a$,
the event horizon is removed 
as $\rho$ no longer approaches to $-\infty$,
On the other hand,
since the perturbation theory does not apply to 
the immediate neighborhood of the horizon at $r = a$,
we cannot rule out the existence of a horizon 
beyond the range of validity of the perturbation theory.

\subsection{Higher Order Corrections}

It is straightforward to calculate the second order corrections 
to $\rho$ and $F$. 
The second order solutions $\rho_2$ and $F_2$ are 
calculated by solving the second order semi-classical Einstein equation and 
expanded around $r=a$ as 
\begin{align}
 \rho_2(r) 
 &= 
 \frac{1}{2(r-a)^2} 
 \left\{a^2 q^2 - \tilde c_1^2 + 2 a \tilde c_1 q \log\left(\frac{r-a}{a}\right)
 - a^2 q^2 \left[\log\left(\frac{r-a}{a}\right)\right]^2\right\} 
 + \mathcal O\left((r-a)^{-1}\right), 
\label{rho2}
\\
 F_2(r) 
 &= 
 \frac{\tilde c_1^2}{a^2} - \frac{2 \tilde c_1 q}{a} 
 + 2 q^2 - \frac{2q}{a} \left(\tilde c_1 - a q\right) \log\left(\frac{r-a}{a}\right) 
 + q^2 \left[\log\left(\frac{r-a}{a}\right)\right]^2 
 + \mathcal O\left((r-a)\right). 
\label{F2}
\end{align}
For $r\sim\mathcal O(\kappa/a)$, 
they are of $\mathcal O(\kappa^{-2})$ and $\mathcal O(\kappa^0 a^{-4})$ respectively. 
Then, the leading, first, and second order terms of the perturbative expansion, 
\eqref{rho-expansion} and \eqref{F-expansion},
have the same order of magnitude, $\mathcal O(\kappa^0 a^0)$ for $\rho$ and 
$\mathcal O(\kappa^2 a^{-4})$ for $F$, respectively. 
This is because the expressions \eqref{rho1}, \eqref{F1}, \eqref{rho2} and \eqref{F2} 
are calculated under the assumption that $r-a\sim\mathcal O(\kappa^0 a)$, 
and hence, the higher order corrections become comparable 
to the lower order terms for $r-a\sim \mathcal O(\kappa/a^2)$. 
This does not imply the breakdown of the perturbative expansion 
and the expansion would be valid if it is calculated 
by using the appropriate assumption, $r-a\sim \mathcal O(\kappa/a^2)$. 
We will discuss this issue again in the subsequent sections.

\subsection{Time-Dependent Perturbations}
\label{time-dependent}

In the above,
we have focused on 
%%% pm %%%
%time-independent perturbations of the metric,
static configurations
%%%
so that the off-diagonal terms $\langle T^t{}_r\rangle$ \eqref{Ttr}
and $\langle T^r{}_t\rangle$ \eqref{Trt} can be set to zeros
by setting $k = 0$.
When $k\neq 0$ in eqs.\eqref{Ttr} and \eqref{Trt},
the energy-momentum tensor is still time-independent 
but we need to consider a time-dependent perturbation of the metric.
We can still use eq.\eqref{metric-1} as the ansatz for the metric,
but $\rho_1$ and $F_1$ are now time-dependent.

It is straightforward to solve the perturbation of the metric
when $k$ is turned on.
The $k$-dependent terms of the energy-momentum tensor are
\begin{align}
\langle T^{t}{}_{t}\rangle &=
\langle T^{r}{}_{r}\rangle = 
\langle T^{\th}{}_{\th}\rangle =
\langle T^{\phi}{}_{\phi}\rangle = 0,
\\
\langle T^{t}{}_{r}\rangle &=
\frac{k}{(r-a)^2}.
\end{align}
The corresponding solution is
\begin{align}
\rho_1 &= \frac{kt}{r-a},
\label{rho1-td}
\\
F_1 &=
\frac{2k(r-a)t}{r^2}.
\label{F1-td}
\end{align}

For the most general time-independent conserved tensor 
\eqref{Ttt}--\eqref{Tphiphi} in the Schwarzschild background,
the solution of the metric perturbation is thus
a superposition of the previous result 
\eqref{rho1-generalT}--\eqref{F1-generalT} 
for time-independent metric
and the time-dependent part \eqref{rho1-td}--\eqref{F1-td}.

%{\color{blue}
%(Is the following statement referring to 
%the perturbative calculation above?)}
%However,
%after the back reaction is taken into consideration at the first order,
%the divergence at the 0-th order is reduced to
%an enhancement by a factor of $\mathcal{O}(a^2/\kappa)$.
%The energy-momentum tensor for $Q\neq 0$
%is now of $\mathcal{O}(1/(\kappa a^2))$.
%(Recall that the Hawking radiation contributes to $\langle T_{uu}\rangle$
%by a term of $\mathcal{O}(1/a^4)$ outside the horizon.)
%Notice that an energy-momentum tensor of $\mathcal{O}(1/(\kappa a^2))$
%is much lower than the Planck scale $\mathcal{O}(1/\kappa^2)$,
%so it is totally sensible to use the low energy effective theory
%in this energy regime.
%The static solutions to the semi-classical Einstein equation
%we have considered so far are perfectly self-consistent.

\section{General Perturbative Solution in Proper Radial Coordinate}
\label{perturbation-z}

In this section,
we repeat the derivation of the metric perturbation 
due to a generic energy-momentum tensor of 4D conformal quantum fields,
but in terms of the proper radial coordinate $z$,
which measures the proper distance along the radial direction.
We will see that certain features of
the near-horizon geometry which is obscured in the $r$-coordinate
is more manifest in the $z$-coordinate.

The ansatz of the metric is eq.\eqref{metric-2}.
The $0$-th order terms are given by the Schwarzschild solution:
\begin{align}
B_0 &= r_0^2(z),
\label{B0}
\\
A_0 &= \log\left[\left(1-\frac{a}{r_0(z)}\right)r_0^4(z)\right],
\label{A0}
\end{align}
where $r_0(z)$ is defined through eq.\eqref{z-r} by
\be
\frac{dr_0(z)}{dz} = \sqrt{1-\frac{a}{r_0(z)}}.
\label{r0-z}
\ee
This equation is invariant under a shift of $z$,
so we can simply choose $z$ such that $z = 0$
at the horizon, i.e. $r_0(z = 0) = a$. 
The function $r_0(z)$ can then be expanded around $z = 0$ as
\be
r_0(z) = a + \frac{z^2}{4a} - \frac{z^4}{48a^3} + \frac{11z^6}{2880a^5}
+ \cdots.
\ee
The 0-th order solutions to $B(z)$ \eqref{B0} and $A(z)$ \eqref{A0}
can also be expanded around $z = 0$ as
\begin{align}
B_0(z) &= a^2 + \frac{z^2}{2} + \frac{z^4}{48a^2} - \frac{z^6}{360a^4}
+ \cdots,
\label{B0-exp}
\\
A_0(z) &= \log\left(\frac{a^2 z^2}{4}\right)
+ \frac{2z^2}{3a^2} - \frac{13z^4}{90a^4} + \cdots.
\label{A0-exp}
\end{align}

To solve the semi-classical Einstein equation \eqref{scEE}
for the perturbative expansion
\begin{align}
B &= B_0 + \kappa B_1 + \cdots,
\\
A &= A_0 + \kappa A_1 + \cdots,
\end{align}
we first solve for $A'_1(z)$ algebraically 
in terms of $B_1(z)$, $\A$ and $f(z)$
from $G^z{}_z = \kappa \langle T^z{}_z\rangle$.
\footnote{
Note that $\langle T^z{}_z\rangle = \langle T^r{}_r\rangle$,
so we can use eq.\eqref{Tupdownrr} to set
\be
\langle T^z{}_z(z)\rangle = \frac{\A}{a(r_0(z)-a)} + f(z)
\label{Tupdownzz}
\ee
for an arbitrary smooth function $f(z)$.
}
One can then use the expression of $A'_1(z)$ to turn
$G^{\mu}{}_{\mu} = \kappa \langle T^{\mu}{}_{\mu}\rangle$
into an equation for $B_1(z)$ only,
and $B_1(z)$ can be solved,
at least as an expansion of $z$.
%YM-6/13
In fact, it would be difficult to solve the first order equations exactly, 
and hence, we will focus on the behavior around $z=0$ and use 
the expansions \eqref{B0-exp} and \eqref{A0-exp}. 
%-6/13

The general solutions to the first order semi-classical Einstein equations are
\begin{align}
B_1 &= 4a^2 \A \log\left(\frac{z}{a}\right)
+ B_{10} 
+ B_{11} \left(\frac{z}{a} + \frac{z^3}{12a^3} + \cdots\right)
+ \frac{a^2}{6}\left[5\A + 3a^2 f(0)\right]
\left(\frac{z}{a}\right)^2 + \cdots,
\label{B1}
\\
A_1 &= 
\frac{B_{11}}{a^2} \left(\frac{2a}{z} + \frac{4z}{3a} + \cdots\right) 
+ 8\A \left[1 - \frac{1}{3}\left(\frac{z}{a}\right)^2\right]
\log\left(\frac{z}{a}\right)
+ A_{10} 
\nn \\
&
+ \frac{1}{9a^2}
\left[-6B_{10} + 4a^2 \A + 18(\b+\c) + 3a^4 f(0)\right]
\left(\frac{z}{a}\right)^2 + \cdots.
\label{A1}
\end{align}
There are three integration constants $A_{10}$, $B_{10}$ and $B_{11}$
in the general solution for the first order perturbation.
$A_{10}$ can always be set to $0$
by rescaling the time coordinate $t$.
$B_{10}$ corresponds to a shift of the Schwarzschild radius $a$,
and thus can be absorbed by a redefinition of $a$.
$B_{11}$ corresponds to a shift of the $z$-coordinate.
In the following, we put $B_{10}=B_{11}=A_{10}=0.$
%Thus one can set all three parameters
%$A_{10}$, $B_{10}$ and $B_{11}$ to $0$'s 
%without loss of generality.

\hide{%%% beginning of hide. %%%
{\color{blue}
The solution above is checked numerically except
terms proportional to $q$.
}
{\color{blue}
(Need to check the correctness of the result above,
which is an incompatible combination of mine with Yoshinori's.
There may be conventional mismatch.)
\begin{align}
 B_1 
 &=  C_0 \left(1 - \frac{1}{48} z^4 + \cdots \right) 
 + C_1 \left(z + \frac{1}{12} z^3 + \cdots\right) 
 \notag\\
 &\quad + C_\ell \left(\log z - \frac{1}{48} z^4 \log z + \cdots 
 + \frac{1}{12} z^2 + \frac{7}{2880}z^4 + \cdots\right)
 \notag\\
 &\quad + (\b+\c)\left(3 z^2 - \frac{1}{2} z^4 + \cdots\right) + \mathcal O(z^5)
\\
 A_1 
 &= C_3  - \frac{2}{3} C_0 \left(z^2 + \cdots\right) 
 + C_1 \left(\frac{2}{z} + \frac{4}{3}z + \cdots\right) 
 \notag\\
 &\quad + C_\ell \left(\frac{1}{18} z^2 + \cdots + 2 \log z 
 - \frac{2}{3} z^2 \log z + \cdots \right) 
 + 4 (\b+\c) \left(z^2 + \cdots\right) + \mathcal O(z^3), 
\end{align}
\begin{align}
C_0 &= B_{10}, \\
C_1 &= B_{11}, \\
C_{\ell} &= 4a^2 q, \\
C_3 &= A_{10}.
\end{align}
}
}%%% end of hide. %%%

Unless $q = 0$,
we have $B_1(z) \rightarrow \pm \infty$ as $z \rightarrow 0$,
and the perturbation theory is valid only
in the region sufficiently far away from the point $z = 0$.
The near-horizon geometry can be classified into 
three categories depending on whether
$q < 0$, $q = 0$ or $q > 0$,
as we have seen in the previous section.
We consider each case separately in the following.

\hide{%%% beginning of hide. %%%
Around the neck,
we have $R^{\mu\nu\lambda\rho}R_{\mu\nu\lambda\rho}$ given by
\be
{\cal F}\simeq{\cal G}\simeq
\frac{12}{a^4} +
\frac{8\kappa}{a^6}\left[
-4a^2\A - 3(2(\b+\c) + B_{10}) + 3a^4 f(0)
- 12a^2\A\log\left(\frac{z}{a}\right)
+ \mathcal{O}(z)
\right] + \mathcal{O}(\kappa^2).
\ee
}%%% end of hide. %%%

\subsection{Wormhole-Like Throat ($q < 0$)}
\label{throat-z}

Let us now focus on the case $q < 0$.
We have seen in Sec.\ref{throat-r}
that there is a wormhole-like throat in this case,
i.e. a local minimum of the areal radius $r$,
or equivalently, a local minimum of $B(z)$.
We will see that the perturbative analysis in the $z$-coordinate
for this case better displays details about the near-horizon geometry.
On the other hand,
the perturbative analysis in the $r$-coordinate in Sec.\ref{throat-r}
%YM-6/13
has the advantage that the first order solutions 
can be obtained without using the expansion in the radial coordinate, 
and hence, the integration constants around $r=a$ can be 
related to those in the asymptotically flat region. 
%has the advantage that the solution is known not only around the neck 
%but also for $r \rightarrow \infty$.
%-6/13

The throat is located
where $B \simeq B_0 + \kappa B_1$ has
a vanishing first derivative.
The condition $B'(z_0) = 0$ implies that
\begin{align}
z \equiv z_0 \simeq 
\sqrt{-4\kappa a^2 q} + \cdots.
%a\sqrt{-\kappa C_{\ell}} + \cdots.
\end{align}
Clearly, the wormhole throat exists only if $q < 0$.

Assuming that $q < 0$,
we expand $B$ and $A$ around the neck at $z = z_0$
and find
\begin{align}
B(z) &\simeq
a^2 
+ \kappa \left[
B_{10} - 2a^2 q + 4a^2 q\log\left(\frac{z_0}{a}\right)
\right]
+ (z-z_0)^2
+ \mathcal{O}(\kappa^{3/2}),
\label{B-pert-z}
\\
A(z) &\simeq
\log(\kappa q)
%+ \frac{2\sqrt{\kappa}B_{11}}{a^2\sqrt{-4 q}}
+ \kappa\left[
%A_{10} 
- \frac{8 q}{3} + 8 q\log\left(\frac{z_0}{a}\right)
\right]
+ \left[
\frac{1}{a\sqrt{-\kappa q}} 
%+ \frac{B_{11}}{2a^3 q}
- \frac{4\sqrt{-\kappa q}}{3a}
\right](z-z_0)
%\nn \\
%&
%+ \left(
%\frac{1}{4\kappa a^3 q}
%+ \frac{2B_{11}\sqrt{-4a^2 q}}{16\kappa a^6 q^2}
%+ \frac{5}{3a^2}
%\right)(z-z_0)^2
%\nn \\
%&
%+ \left(
%\frac{2\sqrt{-4a^2 q}}{48\kappa^{3/2}a^7 q^2}
%- \frac{2B_{11}}{4\kappa a^7 q^2}
%+ \frac{2\sqrt{-4a^2 q}}{12\sqrt{\kappa}a^5 q}
%\right)(z-z_0)^3
+ \mathcal{O}(\kappa^{3/2}),
\label{A-pert-z}
\end{align}
for the neighborhood of $z=z_0$ in which 
\begin{align}
|z-z_0| \lesssim \mathcal{O}(\kappa^{1/2}).
\label{z-z0}
\end{align}
The expansion of $A(z)$ is singular in the limit $\kappa a^2 q \rightarrow 0$
unless the constraint \eqref{z-z0} is imposed.

The second term of $B(z)$ \eqref{B-pert-z}
can always be absorbed in the first term $a^2$
by a redefinition of $a$.
The expansion of $B(z)$ in powers of $(z-z_0)$
is free of the linear term, 
as $z_0$ is chosen to be the point of local extremum.
Note that the coefficient of the quadratic term $(z-z_0)^2$
is always $1$ (up to $\mathcal{O}(\kappa)$ correction)
independent of all parameters.
We will see below that this feature persists
in the non-perturbative analysis.

The first three terms (the constant part) of $A(z)$
can always be set to $0$ by a scaling of the time-coordinate $t$.
\hide{
{\color{blue}
(What about the linear, quadratic and cubic terms of $(z-z_0)$?
They are all determined by only two parameters $C_1$ and $C_{\ell}$.)
}
}
Note that the conformal charges $\b$, $\c$
do not appear in eq.\eqref{B-pert-z} nor \eqref{A-pert-z}.
In terms of the radial distance coordiante $z$,
the conformal anomaly has little effect on the near-horizon geometry
when $q < 0$.

\subsection{Event Horizon ($q = 0$)}
\label{horizon-z}

For the case $q = 0$,
the solution is approximately
\begin{align}
B &=
a^2 + \frac{z^2}{2}
%+ \cdots
%+ \kappa \left[
%4a^2 \A \log\left(\frac{z}{a}\right) + B_{10} + B_{11} \frac{z}{a}
%+ \frac{a^2}{6}\left[5\A + 3a^2 f(0)\right]
%\left(\frac{z}{a}\right)^2
%+ \cdots
%\right] 
+ \kappa \frac{a^2}{2}f(0)z^2
+ \cdots,
\label{B-pert-z-horizon}
\\
A &= 2\log\left(\frac{az}{2}\right)
+ \frac{2z^2}{3a^2} 
%+ \cdots
%+ \kappa \left[
%\frac{B_{11}}{a^2} \left(\frac{2a}{z} + \frac{4z}{3a} + \cdots\right) 
%+ 8\A \log\left(\frac{z}{a}\right)
%+ A_{10} + \cdots \right]
+ \kappa \left[\frac{2(\b+\c)}{a^4} + \frac{f(0)}{3}\right] z^2
+ \cdots
%\nn \\
%&
%+ \frac{1}{9a^2}
%\left[-6B_{10} + 4a^2 \A + 18(\b+\c) + 3a^4 f(0)\right]
%\left(\frac{z}{a}\right)^2 + \cdots.
\end{align}
according to eqs.\eqref{B0-exp}, \eqref{A0-exp}, \eqref{B1} and \eqref{A1}.
Since the first order correction of $\mathcal{O}(\kappa)$
starts only at $\mathcal{O}(z^2)$,
so the near-horizon geometry is only slightly modified
in a small neighborhood of $z = 0$.
It is essentially the same as the Schwarzschild spacetime.
This is in agreement with the result in Sec.\ref{horizon-r}
which was obtained in terms of the $r$-coordinate.

\subsection{No Neck \& No Horizon ($q > 0$)}
\label{nono-z}

For $q > 0$,
we derive from
eqs.\eqref{B0-exp}, \eqref{A0-exp}, \eqref{B1} and \eqref{A1}
the first order perturbation near the horizon:
\begin{align}
B(z) &\simeq
a^2 + \frac{z^2}{2} 
+ 4\kappa a^2 q \log\left(\frac{z}{a}\right)
%+ B_{10} + B_{11} \frac{z}{a}
+ \cdots,
\\
A(z) &\simeq
2\log\left(\frac{az}{2}\right)
+ \frac{2z^2}{3a^2}
+ 8\kappa\A \log\left(\frac{z}{a}\right)
%+ \kappa\left[
%\frac{B_{11}}{a^2} \left(\frac{2a}{z} + \frac{4z}{3a} + \cdots\right) +
%8\A \log\left(\frac{z}{a}\right)
%+ A_{10}
%\right]
+ \cdots.
\end{align}
According to these expressions,
the areal radius has no local minimum and
it vanishes somewhere around $z \sim ae^{-1/(4\kappa q)}$,
where $A(z) \sim \mathcal{O}(1/(\kappa q))$
%%% pm %%%
and $B(z) \sim 0$ for $a^2 \gg \kappa$.
%%%
Strictly speaking,
the perturbation theory is valid only for
\be
z \gg ae^{-1/(4\kappa q)},
\ee
but it shows that the horizon can only exist,
if it does exist,
in the microscopic range of the areal radius.

%%% pm %%%
What we have seen in this section is that,
%%%
in the large-scale geometry,
there is no horizon for both $q < 0$ and $q > 0$.
The horizon exists only when $q$ 
is fine-tuned to exactly $0$.
This result is also compatible with 
the calculation in the $r$-coordinate in Sec.\ref{nono-r}.

\subsection{Higher Order Corrections}

We have seen that the geometry near the horizon 
is modified by perturbative corrections. 
This happens because the perturbative expansion 
around the Schwarzschild solution 
is calculated by assuming $r-a\sim\mathcal
%%% pm %%%
%O(1)$,
O(\kappa^0)$,
%%%
or equivalently, $z\sim \mathcal 
%%% pm %%%
O(\kappa^0)$,
%O(1)$,
%%% 
and hence,
higher order terms can be comparable to lower order terms 
if $z$ is sufficiently small as $z^2 \sim \mathcal O(\kappa)$. 
%Here, we will show that the second order corrections to \eqref{B0-exp}-\eqref{A1} 
%does not modify the structure around the neck \eqref{B-pert-z} and \eqref{A-pert-z}. 
%%% pm %%%
More explicitly,
perturbative expansion
%%%
of $A(z)$ and $B(z)$ are expressed as 
\begin{align}
 A(z) 
 &= 
 \log \left(\frac{z^2}{4 a^2}\right) 
 + \frac{2 z^2}{3 a^2} + 8 \kappa q \log\left(\frac{z}{a}\right) + \cdots, 
\label{A-pert-z-full}
\\
 B(z) 
 &= 
 a^2 + \frac{z^2}{2} + 4 a^2 \kappa q \log \left(\frac{z}{a}\right) + \cdots. 
\label{B-pert-z-full}
\end{align}
The first and second terms of the expressions above come from the 0-th order terms 
of the perturbative expansion and the third terms are first-order corrections. 
If $z$ is small as $z^2 \sim \mathcal O(\kappa)$, 
only the first terms give $\mathcal 
%%% pm %%%
%O(1)$
O(\kappa^0)$
%%%
contributions but 
both the second and third terms become of $\mathcal O(\kappa)$. 
%%% pm %%%
For $q < 0$,
%%%
the local minimum of $B(z)$ moves to the outside of the horizon 
because the second term and third term in eq.\eqref{B-pert-z-full}
are comparable in this region. 
The structure around the local minimum might be modified if 
the higher order terms in eq.\eqref{B-pert-z-full} become comparable to these terms 
when $z^2 \sim\mathcal O(\kappa)$. 
In the following,
we will calculate the second order corrections and 
show that they do not modify the structure of the neck at leading order. 

We consider the perturbative expansion to the second order 
\begin{align}
 B &= B_0 + \kappa B_1 + \kappa^2 B_2 +\cdots, 
\\
 A &= A_0 + \kappa A_1 + \kappa^2 A_2 + \cdots.
\end{align}
The general solutions for $B_2$ and $A_2$ are calculated from 
the second order terms of the semi-classical Einstein equation as 
\begin{align}
 B_2 
 &= 
 B_{20} - 4 a^2 q^2 \log z + B_{21} \frac{z}{a} + \mathcal O(z^2), 
\\
 A_2 
 &= 
 \frac{B_{21}}{a^2}\left(\frac{2}{z} + \frac{4z}{3a}\right) 
 - 16 q^2 \left(\log z\right)^2 + A_{20} + \mathcal O(z^2). 
\end{align}
The integration constants $A_{20}$, $B_{20}$ and $B_{21}$ 
can be set to zero by rescaling the time coordinate $t$, 
shifting the Schwarzschild radius $a$ and shifting 
the $z$-coordinates, respectively, as for those at the first order. 
Then, for $z^2\sim\mathcal O(\kappa)$, 
the above expressions have only $\mathcal 
%%% pm %%%
%O(1)$
O(\kappa^0)$
%%%
terms, which 
give contributions of $\mathcal O(\kappa^2)$ to $B$ and $A$. 
Therefore, the expressions \eqref{A-pert-z-full} and \eqref{B-pert-z-full} are sufficient to determine the behavior of $A$ and $B$ up to $\mathcal O(\kappa^2)$. 

In order to see why the perturbative expansion for $\rho$ and $F$ 
in terms of $r$-coordinate is not good for $r-a \sim \mathcal O(\kappa)$, 
we write the proper radial coordinate $z$ as a function of $r$. 
From eq.\eqref{B-pert-z-full}, $z^2$ is expressed as 
\begin{equation}
 z^2 = 2 (r^2-a^2) - 4 a^2 \kappa q \log\left(\frac{r^2-a^2}{a^2}\right) 
 - \frac{16 a^4 \kappa^2 q^2}{r^2-a^2} \log\left(\frac{r^2-a^2}{a^2}\right) 
 + \mathcal O(\kappa^3), 
\label{z-exp-r}
\end{equation}
assuming that $r-a\sim \mathcal 
%%% pm %%%
%O(1)$.
O(\kappa^0)$.
%%% 
As the expression above is calculated from the first 3 terms in eq.\eqref{B-pert-z-full}, 
which give $\mathcal O(\kappa)$ contributions for $z^2\sim\mathcal O(\kappa)$, 
all terms in eq.\eqref{z-exp-r}, including higher order terms, 
become $\mathcal O(\kappa)$ and comparable. 
This implies that the perturbative expansion in terms of $r$ 
(which is calculated assuming $r-a\sim\mathcal 
%%% pm %%%
%O(1)$
O(\kappa^0)$)
%%%
does not give a good expansion for $r-a\sim\mathcal O(\kappa)$, 
and higher order terms give contributions of the same order.

\section{Perturbative Solution Near Schwarzschild Radius}
\label{near-Schwarzschild}

In the previous sections, we have studied 
perturbations around the Schwarzschild solution. 
In general, the first-order correction becomes very large 
near the Schwarzschild radius $r=a$, 
implying that the perturbative expansion around the Schwarzschild solution 
is not good near the Schwarzschild radius. 
However, the perturbative expansion there is 
simply the small-$\kappa$ expansion for the semi-classical Einstein equation. 
Assuming that the semi-classical Einstein equation has a solution for arbitrary $\kappa$, 
the perturbative expansion should be possible at arbitrary $r$. 
This implies that the zero-th order solution in the perturbative expansion is 
not given by the Schwarzschild solution near the Schwarzschild radius. 
In this section, we focus on the solution near the Schwarzschild radius 
and study the perturbative expansion there. 

The perturbative expansion around the Schwarzschild solution 
is valid for 
\begin{equation}
 r - a \gg \frac{\kappa}{a}, 
 \label{OutNearSS}
\end{equation}
but it is not good if $r$ is very close to the Schwarzschild radius as 
\begin{equation}
 r \sim a + \frac{\kappa}{a},
 \label{NearSS}
\end{equation}
and hence we consider the perturbative expansion in this region. 
In order to focus on this region, 
we introduce the tortoise coordinate $x$ as 
\begin{equation}
 dx^2 = \frac{dr^2}{F(r)} = e^{-\rho(x)} dz^2,
 \label{tortoise}
\end{equation}
and then, the metric is expressed as 
\begin{equation}
ds^2 = - e^{\rho(x)} \left[dt^2 - dx^2\right] + r^2(x) d\Omega^2.
\end{equation}
We focus on the geometry near $r=a$. 
In eq.\eqref{OutNearSS}, where the expansion around the Schwarzschild solution is good, 
$e^{\rho(x)}$ is given by 
\begin{equation}
 e^{\rho(x)} = 1 - \frac{a}{r(x)}, 
\end{equation}
and approaches to $\mathcal O(\kappa)$ as $r$ goes to $r = a + \mathcal O(\kappa)$. 
For the junction condition for the expansion in \eqref{OutNearSS} and that in \eqref{NearSS}, 
$e^{\rho(x)}$ must behave as 
\begin{equation}
 e^{\rho(x)} = \mathcal O(\kappa). 
\end{equation}
Therefore, in \eqref{NearSS}, 
$\rho(x)$ and $r(x)$ should be expanded as 
\begin{align}
 e^{\rho(x)} 
 &= 
 \kappa e^{\tilde\rho_0(x) + \kappa\rho_1(x)} + \cdots, 
\label{rho-nss}
\\
 r(x) 
 &= 
 a + \kappa r_1(x) + \cdots. 
\label{r-nss}
\end{align}

We solve the semi-classical Einstein equation, \eqref{Eq-trace} and \eqref{Eq-rr}, 
in the expansion above to first order, $\rho_1(x)$ and $r_1(x)$. 
The trace part of the energy-momentum tensor is 
given by the trace anomaly \eqref{anomaly}, 
while $\langle T^x{}_x\rangle$ is given by \eqref{Tyy}. 
Near the Schwarzschild radius, 
they are expressed in terms of the expansion above as 
\begin{align}
 \langle T^\mu{}_\mu \rangle 
 &= 
 \frac{c_4}{3\kappa^2} e^{-2 \tilde\rho_0(x)} 
 \tilde\rho_0^{\prime\prime\,2}(x) + \mathcal O(\kappa^{-1}), 
\\
 \langle T^x{}_x\rangle 
 &= 
 \frac{q e^{-\tilde\rho_0(x)}}{\kappa a^2} + \mathcal O(\kappa^0). 
\end{align}
Here, we assumed that $\langle T^\theta{}_\theta\rangle$ does not diverge 
as $e^{\rho(x)} \to 0$, and hence, $H(x)$ and $G(x)$ are of $\mathcal O(\kappa)$. 

The leading order term of \eqref{Eq-trace}, which is of $\mathcal O(\kappa^{-1})$, 
is calculated as 
\begin{equation}
 e^{- \tilde\rho_0(x)} \tilde\rho''_0(x) 
 = \frac{c_4}{3} e^{-2 \tilde\rho_0(x)} \tilde\rho_0^{\prime\prime\,2}(x), 
\end{equation}
and a solution is given by 
\begin{equation}
 e^{\tilde \rho_0(x)} = \tilde c_0 e^{\lambda_0 x},
 \label{rho0-x}
\end{equation}
where $c_0$ and $\lambda_0$ are integration constants. 
There is another solution which is the solution of 
\begin{equation}
 \tilde\rho''_0(x) - c_4 e^{\tilde\rho_0(x)}
 = 0, 
\end{equation}
which gives large curvature of $\mathcal O(\kappa^{-1})$, 
and should be excluded as an unphysical solution. 

By substituting the first order solution \eqref{rho0-x}, 
the next-to-leading order terms of the semi-classical Einstein equation 
\eqref{Eq-trace} and \eqref{Eq-rr} give the following differential equations 
for $\rho_1(x)$ and $r_1(x)$:
\begin{align}
 0 &= \tilde c_0 e^{\lambda_0 x} + q - a \lambda_0 r_1'(x), 
\\
 0 &= - 2 \tilde c_0 e^{\lambda_0 x} + a^2 \rho''_1(x) + 4 a r''_1(x), 
\end{align}
and they are solved as 
\begin{align}
 \rho_1(x) 
 &= 
 \bar \rho_1 + \lambda_1 x - \frac{2 \tilde c_0 e^{\lambda_0 x}}{a^2 \lambda_0^2}, 
% - \int dx \frac{\bar C_0 e^{\lambda_0 x}}{\lambda_0} f(x), 
\\
 r_1(x) 
 &= 
 a_1 + \frac{\tilde c_0 e^{\lambda_0 x} + \lambda_0 q x}{a \lambda_0^2}, 
% + a \int dx \frac{\bar C_0 e^{\lambda_0 x}}{\lambda_0} f(x),  
\end{align}
where $a_1$, $\bar \rho_1$ and $\lambda_1$ are integration constants, 
which can be absorbed by redefinitions of $a$, $\tilde c_0$ and $\lambda_0$, respectively. 
The solution to the first order perturbation is obtained as 
\begin{align}
 e^{\rho(x)} 
 &= 
 \kappa \tilde c_0 e^{\lambda_0 x}
 \left(1 - \frac{2 \kappa \tilde c_0 e^{\lambda_0 x}}{a^2 \lambda_0^2} \right) 
 + \mathcal O(\kappa^3), 
\label{sol-rho-x}
\\
 r(x) 
 &= 
 a + \kappa \frac{\tilde c_0 e^{\lambda_0 x} + \lambda_0 q x}{a \lambda_0^2} 
 + \mathcal O(\kappa^2).
\label{sol-r-x}
\end{align}

\subsection{Wormhole-Like Throat ($q<0$)}

If $q<0$, $r$ has a local minimum as we have seen in the previous sections. 
From eq.\eqref{sol-r-x}, the local minimum is located at 
\begin{equation}
 x = x_0 \equiv \frac{1}{\lambda_0} \log \frac{|q|}{\tilde c_0}, 
\end{equation}
and $r$ is expanded as 
\begin{equation}
 r(x) 
 \simeq 
 a + \frac{\kappa q}{a \lambda_0^2} \left(-1 + \log \frac{|q|}{\tilde c_0}\right) 
 + \frac{\kappa |q|}{2 a} \left(x- x_0\right)^2 + \cdots. 
\end{equation}
In terms of the proper radial coordinate, 
$B(z) = r^2(z)$ is expanded as 
\begin{equation}
 B(z) \simeq a^2 + \left(z-z_0\right)^2 + \cdots, 
\label{B-x}
\end{equation}
where we redefined $a$ again to absorb
the $\mathcal O(\kappa)$ constant term above. 
This is consistent with the result \eqref{B-pert-z} in the previous section.

\subsection{Event Horizon ($q=0$)}

For $q=0$, $r$ is given by 
\begin{equation}
 r(x) = a + \frac{\kappa \tilde c_0 e^{\lambda_0 x}}{a \lambda_0^2} + \mathcal O(\kappa^2). 
\end{equation}
In this case, $r$ does not have local minimum but approaches to $r=a$ as $x\to - \infty$. 
The event horizon is located in the limit $x\to-\infty$
in the tortoise coordinate for classical solution 
and hence this behavior is consistent with the event horizon. 
In the proper radial coordinate, 
$B(z) = r^2(z)$ is expanded as 
\begin{equation}
 B(z) = a^2 + \frac{1}{2} z^2 + \cdots, 
\label{B-x-horizon}
\end{equation}
which is also consistent with the result in the previous section, \eqref{B-pert-z-horizon}.

%%% pm %%%
%\subsection{No Neck or Horizon}
\subsection{No Neck \& No Horizon ($q > 0$)}
%%%

For $q>0$, \eqref{sol-r-x} is a monotonic increasing function 
and has no local minimum. 
The areal radius $r$ can be smaller than $a$ but 
$e^{\rho(x)}$, whose leading order term is given by \eqref{rho0-x}, 
approaches to zero but never goes to zero, 
at least in this near-Schwarzschild-radius region. 
Therefore, there are no neck or event horizon. 
In terms of the proper radial coordinate $z$, 
$B(z)$ is expressed as 
\begin{equation}
 B(x) = a^2 + \frac{1}{2} z^2 + \frac{4\kappa q}{\lambda_0^2} \log \frac{\lambda_0 z}{2} 
 + \mathcal O(\kappa^2). 
\end{equation}
This is consistent with the result in the previous section 
if $\lambda_0 \sim a^{-1}$.

\subsection{Relation to Expansion Around Schwarzschild Metric}

In the previous section, we have calculated the perturbative expansion 
around the Schwarzschild solution assuming that $r-a\sim\mathcal 
%%% pm %%%
%O(1)$,
O(\kappa^0)$,
%%% 
and shown that the expression in terms of the proper radial coordinate $z$ 
gives a good expansion even for $r-a\sim\mathcal O(\kappa)$. 
The second order corrections do not contribute to 
the leading order terms, which is of $\mathcal O(\kappa)$ 
of the perturbative expansion for $r-a\sim\mathcal O(\kappa)$, 
and it is expected that the higher order terms of the expansion for $r-a\sim\mathcal 
%%% pm %%%
%O(1)$
O(\kappa^0)$
%%% 
would not affect the leading order terms of the expansion for $r-a\sim\mathcal O(\kappa)$.

The perturbative expansion for $r-a\sim\mathcal 
%%% pm %%%
%O(1)$
O(\kappa^0)$
%%%
is expressed in terms of $z$ as 
eqs.\eqref{A-pert-z-full} and \eqref{B-pert-z-full}. 
The expansions of $\rho$ and $r$ are  
\begin{align}
 \rho(z) 
 &= 
 \log\left(\lambda^2 a^2\right) + \log\left(\frac{z^2}{4 a^2}\right) - \frac{z^2}{3a^2} + \cdots, 
\\
 r(z) 
 &= 
 a + \frac{z^2}{4 a} + 2 a \kappa q \log\left(\frac{z}{a}\right) + \cdots. 
\end{align}
Here, we have introduced an additional constant $\lambda$
which can be absorbed 
by a redefinition of the time coordinate $t$. 
The proper radial coordinate $z$ is related to
the tortoise coordinate $x$ by \eqref{tortoise}, 
and the relation can be expressed as
\begin{equation}
 x - x_0 \simeq \frac{2}{\lambda} \log\left(\frac{z}{a}\right) + \frac{z^2}{6 a^2 \lambda}. 
\end{equation}
For small $z$ as $z^2\sim\mathcal O(\kappa)$, 
the constant $x_0$ should behave as $x_0\sim\log\kappa$, 
and hence we write 
\begin{equation}
 x_0 \equiv - \frac{1}{2\lambda} \log\left(4 \kappa \hat c_0\right). 
\end{equation}
Then, $z$ can be expressed as 
\begin{equation}
 \frac{z}{a} 
 = 
 2 (\kappa \hat c_0)^{1/2} e^{\lambda x/2} \left(1 - \frac{\kappa \hat c_0}{3} e^{\lambda x}\right). 
\end{equation}
As a result, $\rho$ and $r$ are rewritten as 
\begin{align}
 e^{\rho(x)}
 &= 
 \kappa \hat c_0 e^{\lambda x}
 \left(1 - 2 \kappa \hat c_0 e^{\lambda x}\right) + \mathcal O(\kappa^3), 
\\
 r(x) 
 &= 
 a + a \kappa q \log(4 \kappa \hat c_0) 
 + \kappa \hat c_0 e^{\lambda x} + a \kappa q \lambda x + \mathcal O(\kappa^2). 
\end{align}
Here, the additional constant $a \kappa q \log(4 \kappa \hat c_0)$ 
can be absorbed by a redefinition of $a$. 

The expressions above agree with 
\eqref{sol-rho-x} and \eqref{sol-r-x} if 
\begin{align}
 \hat c_0 &= \tilde c_0,
\\
 \lambda &= \lambda_0 = \frac{1}{a}. 
\end{align}
This implies that the perturbative expansions for $z^2\sim\mathcal 
%%% pm %%%
%O(1)$
O(\kappa^0)$
%%%
(eqs.\eqref{A-pert-z-full} and \eqref{B-pert-z-full})
completely reproduce the leading order terms of the 
perturbative expansions for $z^2\sim\mathcal O(\kappa)$, 
and hence the higher order corrections to 
eqs.\eqref{A-pert-z-full} and \eqref{B-pert-z-full} 
do not contribute to the leading order terms of
the perturbative expansion around the neck. 
The constant $\lambda$ is fixed here 
since a boundary condition is imposed
on the expansion around the Schwarzschild metric.

\section{Non-Perturbative Analysis}
\label{nonpert-analysis}
In the perturbative analysis,
we have seen that the near-horizon region can 
be classified into the following three categories:
(1) wormhole-like throat:
There is a local minimum of the areal radius,
resembling the throat of a traversable wormhole,
(2) black-hole horizon:
It is essentially the Schwarzschild event horizon.
(3) neither of the above.
In this section,
we shall find non-perturbative solutions to
the semi-classical Einstein equations.
They are also classified into three categories 
that are compatible with the perturbative result,
while the non-perturbative solutions also demonstrate
new features that are not shown in the perturbative analysis.

The metric that we will use for the non-perturbative analysis
is eq.\eqref{metric-2},
which describes a generic static, spherically symmetric geometry.
For the reader's convenience, 
it is repeated here:
\be
ds^2 = - \frac{e^{A(z)}}{B^2(z)} dt^2 + dz^2 + B(z) d\Omega^2.
\label{metric-z}
\ee
We shall solve the functions $A(z)$ and $B(z)$ 
in a neighborhood of an arbitrary point
from the semi-classical Einstein equation.

As we wish to consider the most general vacuum energy-momentum tensor
for a conformal matter field,
the only constraints on the vacuum energy-momentum tensor 
are (i) the anomaly condition \eqref{anomaly}
and (ii) the conservation law \eqref{conservation}.
Similar to the perturbative analysis in Sec.\eqref{perturbation-z},
our strategy is to first write down
the most general conserved energy-momentum tensor
$\langle T^{\mu}{}_{\nu} \rangle$
compatible with these two constraints,
and then solve two independent semi-classical Einstein equations,
e.g.
\begin{align}
G^z{}_z = \kappa \langle T^z{}_z \rangle
\qquad \mbox{and} \qquad
G^{\mu}{}_{\mu} = \kappa \langle T^{\mu}{}_{\mu} \rangle.
\label{nonpert-EE}
\end{align}
While the trace $\langle T^{\mu}{}_{\mu} \rangle$ is fixed
by the anomaly \eqref{anomaly},
the energy-momentum is now uniquely determined by
$\langle T^z{}_z \rangle$. 
Equivalently, 
it is fixed by a constant parameter $q$
together with a functional degree of freedom
that can be attributed to $\Theta$
or $\langle T^{\th}{}_{\th} \rangle$, as in \eqref{Tyy} 
with \eqref{H-def-nonpert} and \eqref{G-def-nonpert}.

With the generic conserved energy-momentum tensor given above,
we only need to solve two of the semi-classical equations \eqref{nonpert-EE},
which are
\be
G^z{}_z = \kappa(B(z)e^{-A(z)}q + f(z))
\label{Eqzz-nonpert}
\ee
and
\be
G^{\mu}{}_{\mu} 
%= \kappa \langle T^{\mu}{}_{\mu}\rangle,
= \kappa (\b{\cal F} + \c{\cal G}),
\label{Eqtrace-nonpert}
\ee
where $f(z)$ is different from \eqref{f-pert} but defined by 
\be
f(z) \equiv B(z)e^{-A(z)}
\left[H(z) + G(z)\right], 
\ee
and  $H(z)$ and $G(z)$ are given by \eqref{H-def-nonpert} and \eqref{G-def-nonpert}.
%It should be noted that $z_0$ in the calculations below is different from 
%that in \eqref{H-def-nonpert} and \eqref{G-def-nonpert}, 
%and hence $f(z_0)$ may not be zero at $z=z_0$. 
%In \eqref{H-def-nonpert} and \eqref{G-def-nonpert}, 
%$z_0$ is chosen such that $H(z_0) = G(z_0) = 0$ when $A(z)\to -\infty$ in $z\to z_0$, 
%but in the calculations below, we will consider expansion around a point, 
%which will be referred to as $z=z_0$. 

%As $H_{z_0}(z_0) = G_{z_0}(z_0) = 0$,
%we have $f(z) \sim \mathcal{O}(z-z_0)$
%as $z \rightarrow z_0$,
%so that the first term in eq.\eqref{Eqzz-nonpert}
%dominates over the second term $f(z)$
%in a sufficiently small neighborhood of $z_0$.

By integrating \eqref{Eqzz-nonpert}, $A(z)$ is expressed in term of $B(z)$ as 
\begin{equation}
 A(z) = A_0 + \int_{z_0}^z K(z') dz' 
 + \log\left\{1 
 + 2 \kappa q \int_{z_0}^z \exp\left[-A_0-\int_{z_0}^{z'} K(z'') dz''\right] 
 \frac{B^2(z')}{B'(z')} dz'\right\}, 
\label{A-in-B}
\end{equation}
where 
\begin{align}
 K(z) 
&= 
 \frac{2}{B'(z)} \left(1 + \kappa f(z) B(z) \right) + \frac{3B'(z)}{2B(z)}, 
\label{K-np}
\end{align}
and $A_0$ is the integration constant. 
Then eq.\eqref{Eqtrace-nonpert} gives the differential equation for $B(z)$. 
It is hard to find exact solutions to
%the semi-classical Einstein equations
%\eqref{Eqzz-nonpert} and 
\eqref{Eqtrace-nonpert},
so we shall solve them in a small neighborhood
around a fixed point $z = z_0$. 
In eq.\eqref{A-in-B},
%%% pm %%%
%it is convenient to define integrations 
%such that the integration becomes zero at $z=z_0$, 
%but it can be put in different point by choosing an appropriate $A_0$. 
we have used an integral from $z_0$ to $z$,
but it can be replaced by a different starting point for integration 
with an appropriate value of $A_0$
%%%
In general,
integrands may diverge at $z=z_0$, 
and then the integration would be defined such that 
it has no constant part in its expansion around $z=z_0$. 

Since we do not intend to attack the UV problem close to the origin,
we shall always assume that the areal radius $r$ is large 
in this neighborhood,
so that the function $B(z)$ (which is $r^2$) is finite
and much larger than $\kappa$.
On the other hand,
the function $A(z)$ can be either finite or diverging.
($A(z) \rightarrow -\infty$ as $z\rightarrow z_0$
if there is a horizon at $z_0$.)
%We shall first find the non-perturbative solution
%around a generic point $z = z_0$ which is not a horizon
%in Sec.\ref{finite-A},
%and then identify the special case when
%it is a local minimum of the areal radius in Sec.\ref{nonpert-wormhole}.
%Then we study in Sec.\eqref{nonpert-horizon}
%the case when $z = z_0$ is the location of an event horizon.

%%% pm %%%
%\subsection{Finite $A(z_0)$}
\subsection{Solving $A(z)$ and $B(z)$}
%%%
\label{finite-A}

In a sufficiently small neighborhood of a generic point $z = z_0$
which is not a horizon,
we can expand the functions $A(z)$ and $B(z)$ in power of $z$ as
\begin{align}
A(z) &= A_0 + A_1 (z-z_0) + A_2 (z-z_0)^2 + A_3 (z-z_0)^3 + \cdots,
\label{A-exp-3}
\\
B(z) &= B_0 + B_1 (z-z_0) + B_2 (z-z_0)^2 + B_3 (z-z_0)^3 + \cdots.
\label{B-exp-3}
\end{align}
Plugging them into the semi-classical Einstein equations
\eqref{Eqzz-nonpert} and \eqref{Eqtrace-nonpert},
we find straightforwardly the solutions
for the coefficients $A_n$, $B_n$:
\begin{align}
A_1 = 
\frac{2}{B_1} + \frac{3B_1}{2B_0}
+ \frac{2\kappa B_0}{B_1} 
\langle T^z{}_z(z_0) \rangle,
\label{A1-sol}
\end{align}
where
$
\langle T^z{}_z(z_0) \rangle = \left[q_{z_0} B_0 e^{-A_0} + f(z_0)\right],
$
and
\begin{align}
A_2 &= - \frac{B_0^2 A_1^2 + 3B_1^2 - 2B_0(2 + A_1 B_1)}{4B_0^2}
+ \mathcal{O}(\kappa),
\label{A2-sol1}
\\
B_2 &= \frac{B_1(-8B_0 + B_0^2 A_1^2 - 3B_1^2)}{4B_0(B_0 A_1 - 3B_1)}
+ \mathcal{O}(\kappa),
\label{B2-sol1}
\\
A_3 &= 
\frac{64B_0^3 + 16B_0^2 B_1^2 - 52B_0 B_1^4 + 27B_1^6}{96B_0^3 B_1^3}
+ \mathcal{O}(\kappa),
\label{A3-sol1}
\\
B_3 &= 
\frac{4B_0B_1 - B_1^3}{48B_0^2}
+ \mathcal{O}(\kappa),
\label{B3-sol1}
\end{align}
etc. for arbitrary non-zero constants $B_0$, $B_1$.

Notice that this solution is not unique.
There is in fact another class of solutions given by
eq.\eqref{A1-sol} and
\begin{align}
A_2 &= 
\frac{3(B_0 A_1 - 3B_1)^2}{2\kappa \b B_0^2 A_1^2} 
+ \mathcal{O}(\kappa^0),
\label{A2-sol2}
\\
B_2 &=
- \frac{3(B_0 A_1 - 3B_1)B_1}{2\kappa \b B_0 A_1^2} 
+ \mathcal{O}(\kappa^0),
\label{B2-sol2}
\\
A_3 &= 
\frac{288B_0^2 B_1^3(4B_0-3B_1^2)^2}{\kappa^2 \b^2 (4B_0+3B_1^2)^5}
+ \mathcal{O}(\kappa^{-1}),
\label{A3-sol2}
\\
B_3 &=
\frac{48B_0^3 B_1^3(4B_0-9B_1^2)(4B_0-3B_1^2)}{\kappa^2 \b^2(4B_0+3B_1^2)^5}
+ \mathcal{O}(\kappa^{-1}),
\label{B3-sol2}
\end{align}
etc.
This solution has a singular limit as $\kappa \rightarrow 0$.
The energy-momentum tenor for this configuration is of the Planck scale,
presumably due to the presence of classical conformal matter,
rather than being that of a vacuum state.
According to the Buchdahl theorem \cite{Buchdahl:1959zz},
any static classical matter configuration with
a radius less than $9/8$ of the Schwarzschild radius
must suffer divergence in pressure.
What we see here is that a regular solution exists
for conformal matter due to quantum effect.
The energy density and pressure are regularized to the Planck scale.
The same effect was shown in Ref.\cite{Ho:2017vgi}
with the 2D model for the energy-momentum tensor.
It was also obtained in a 4D model \cite{Kawai:2013mda,Kawai:2014afa,Ho:2015fja,Kawai:2015uya,Ho:2015vga,Ho:2016acf,Kawai:2017txu}.

We shall focus on the solution \eqref{A1-sol}--\eqref{B3-sol1}
in the following.

Notice also that eq.\eqref{A1-sol} is singular if
either $B_0$ or $B_1$ vanishes.
While $B_0 \gg \kappa$ as it is the areal radius squared at $z = z_0$,
it is possible that $B_1$ vanishes at $z = z_0$,
which means that it is a local minimum of the areal radius.

For the case $B_1 \neq 0$,
$A_0$ is arbitrary,
and all other coefficents are fixed for given $A_0$, $B_0$ and $B_1$
as in eqs.\eqref{A1-sol}--\eqref{B3-sol1}.
For the solutions in which there is neither a horizon 
nor a wormhole-like structure,
the geometry is given by this solution 
everywhere in in vacuum
as long as $B_0$ is sufficiently large.

On the other hand,
even if $B_1 \neq 0$ at $z = z_0$,
there may be another point $z = z_0'$ where $B_1$ vanishes
when $B(z)$ is expanded around $z_0'$.

There are thus three classes of solutions.
One class of solutions has $A(z)$ diverging at a certain point $z = z_0$,
to be discussed in Sec.\ref{nonpert-horizon}.
For the other two classes,
$A(z)$ is finite everywhere,
but $B(z)$ either has or has no local minimum.
We will focus on the class of solutions with 
a local minimum of $B(z)$ in Sec.\ref{nonpert-wormhole}.

Regardless of whether $B_1$ vanishes,
the tangential pressure at the point $z = z_0$ is
%\begin{align}
%\langle T_{\th\th}(0)\rangle =
%\frac{3B_0(4B_0-B_1^2)(4B_0-3B_1^2)}{2\kappa b(4B_0+3B_1^2)^2}
%\qquad \mbox{or} \qquad
$\langle T_{\th\th}(0)\rangle =
\mathcal{O}(\kappa)$
%\end{align}
for the solution \eqref{A1-sol}--\eqref{B3-sol1}.
(Incidentally,
$\langle T_{\th\th}(0)\rangle =
\mathcal{O}(\kappa^{-1})$
for the solution eqs.\eqref{A2-sol2}--\eqref{B3-sol2}.)

%{\color{blue}
%What are $\langle T^t{}_t \rangle$, $\langle T^z{}_z \rangle$
%and $\langle T^{\th}{}_{\th} \rangle$?
%Maybe we should not dismiss the solutions 
%which diverge in the limit $\kappa \rightarrow 0$.
%}

\subsection{Wormhole-Like Throat ($q<0$)}
\label{nonpert-wormhole}

In this subsection,
we focus on the nonperturbative solution to 
the semi-classical Einstein equations
that has a wormhole-like structure,
that is,
there is a local minimum of the areal radius $r$,
and we study the solution
in a small neighborhood of the wormhole neck. 

%%% pm %%%
%Let the spacetime has something non-trivial at $r=a$. 
%%%
If $r=a$ is located in finite proper distance from finite $r$ with $r\neq a$, 
$B(z)$ should have a regular expansion around $r=a$, as \eqref{B-exp-3}:
\begin{equation}
 B(z) = a^2 + B_1 \left(z-z_0\right) + B_2 \left(z-z_0\right)^2 + B_3 \left(z-z_0\right)^3 + \cdots. 
\end{equation}
%Here, we shall refer to the areal radius of the wormhole neck 
%as the quantum Schwarzschild radius and denote it by $r = a$.
%As a result,
%$B_0 = a^2$.
%%% pm %%%
%In order for $A(z)$ to have a non-trivial expansion at $z=z_0$, 
%$B'(z)$ should go to zero there, or equivalently, $B_1=0$. 
Let us assume that $B_1 = 0$,
%%%
then $z=z_0$ is the local minimum of $r$,
and $B(z)$ is expanded as 
\begin{equation}
 B(z) = a^2 + B_2 \left(z-z_0\right)^2 + B_3 \left(z-z_0\right)^3 + \cdots, \label{B-exp-np}
\end{equation}
with $B_2$ positive.%
\footnote{%
To be more precise, 
$B_2$ can be negative in general, 
%%% pm %%%
%but the first zero of $B_1$ from outside must have $B_2\geq 0$,
but $B_2$ must be non-negative 
at the first zero of $B_1$ as one moves towards the center from distance,
%%%
since $B(z)$ is an increasing function in the asymptotic region. 
The argument here (that $A(z)$ does not diverge) 
can be generalized to $B_2=0$ with $B_n>0$ for 
the first non-zero $B_n$ with $n>0$. 
} 
Here, $a$ is nothing but the areal radius of the wormhole neck, 
to which we shall refer as the quantum Schwarzschild radius. 
Then $K(z)$, defined by eq.\eqref{K-np}, behaves as 
\begin{equation}
 K(z) = \frac{1 + a^2 \kappa f(z_0)}{B_2} \frac{1}{z} + \text{(finite)}, 
\end{equation}
and hence, the second term in \eqref{A-in-B} 
has a logarithmic divergence in the limit $z\to z_0$. 
For $q\neq 0$, 
the third term in \eqref{A-in-B} behaves as 
\begin{equation}
 \log\left[-e^{-A_0}\frac{\kappa q a^4}{1+ a^2 \kappa f(z_0)} z^{-(1+a^2 f(z_0))/B_2} + \cdots \right], 
\end{equation}
and hence, the logarithmic divergences 
in the second and third terms in \eqref{A-in-B} cancel. 
The integration constant $A_0$ in the expression above is also cancelled 
with the first term in \eqref{A-in-B}, and $A(z)$ is then expanded as 
\begin{equation}
 A(z) = \log\left[-\frac{\kappa q a^4}{1+ a^2 \kappa f(z_0)}\right] + \cdots, 
\end{equation}
in which the leading order term is real only if $q<0$. 
This implies that the assumption \eqref{B-exp-np}, 
or equivalently, the local minimum of $r$ appears only if $q<0$. 
This is consistent with the results in the previous sections.  

Now, we have seen that $A(z)$ and $B(z)$ do not diverge for $q<0$ and 
have regular expansion as \eqref{A-exp-3} and \eqref{B-exp-3}. 
%Next, we consider the expansion \eqref{A-exp-3} and \eqref{B-exp-3}. 
Next, we consider \eqref{A-exp-3} and \eqref{B-exp-3} in more details 
and calculate the coefficient. 
Assuming that the local minimum is located at $z = z_0$,
so the expansion of $B(z)$ \eqref{B-exp-3} around this point
has $B_1 = 0$.
Then eq.\eqref{A1-sol} implies that
$A_1$ remains arbitrary,
but $A_0$ and $B_0$ must satisfy the relation
\be
\langle T^z{}_z(z_0) \rangle 
= q_{z_0} B_0 e^{-A_0} + f(z_0) = - \frac{1}{\kappa B_0}.
%B_1 = 0 \qquad \Rightarrow \qquad
%1 + \kappa B_0\left[q B_0 e^{-A_0} + f(0)\right] = 0.
\ee
All other coefficients are fixed for given $A_0$ (or $B_0$) and $A_1$.
There are only 2 free parameters, 
instead of 3 as the case $B_1 \neq 0$ 
simply because the location $z_0$ of the local minimum of $B(z)$
is another free parameter.

%We shall refer to the areal radius of the wormhole neck 
%as the quantum Schwarzschild radius and denote it by $r = a$.
%As a result,
%$B_0 = a^2$.

%Assuming that there is a neck
%(a local minimum of the areal radius $r$),
%we focus on a small neighborhood of the neck
%with the ansatz \eqref{metric-2} with
%\begin{align}
%A(z) &= 2 (H_0 + H_1 z + H_2 z^2 + H_3 z^3 + \cdots),
%\\
%B(z) &= a^2 + B_2 z^2 + B_3 z^3 + \cdots,
%\end{align}
%where we have shifted the coordinate $z$
%such that $B_1 = 0$.
%We can also scale the time coordinate such that $H_0 = 0$.

%We can then solve the 2 Einstein equations 
%\eqref{Eq-trace}, \eqref{Eq-thth}
%around the neck in the $z$-expansion.
%First, we solve the two equations
%\eqref{Eq-trace}, \eqref{Eq-thth} at the 0-th power of $z$.
%The solutions are

Let us go through the derivation of the solutions
\eqref{A1-sol}--\eqref{B3-sol1} and \eqref{A2-sol2}--\eqref{B3-sol2}
for the special case of the wormhole-like structure
in more detail.
First, 
while $A_1$ is a free parameter,
$A_2$ and $B_2$ are solved from the semi-classical Einstein equation as 
%%% Y = 2X %%%
\begin{align}
A_2 &= Y - \frac{A_1^2}{4},
\\
B_2 &= 
a^2 Y
%2a^2 X 
- \kappa \langle T_{\th\th}(z_0)\rangle,
\end{align}
where
\begin{align}
Y &\equiv
\left(16 \kappa a^4 \b\right)^{-1}
\left\{
3a^4 + 4a^2\kappa[\b(6\kappa\langle T_{\th\th}(z_0)\rangle-2)-3\c]
\right.
\nn \\
&
\left.
\pm \sqrt{9a^8 + 72\kappa a^6(2\kappa\b\langle T_{\th\th}(z_0)\rangle-2\b-\c)
+48a^4\c\kappa^2 (4\kappa\b\langle T_{\th\th}(z_0)\rangle+4\b+3\c)} \right\}.
\label{Y-sol}
\end{align}
%\be
%Y \equiv
%\frac{A_1^2}{4} + A_2.
%%X \equiv \frac{H_1^2}{2} + H_2.
%\label{XH1H2}
%\ee
%Note that the solution \eqref{X-sol} of $X$
%is inversely proportional to the anomaly charge $\kappa b$.
%This shows the non-perturbative nature of the solution.
%It is interesting that $B_2$ is exactly fixed by
%the anomaly and the value of $\langle T_{\th\th}\rangle $ at the neck,
%but is independent of the functional form of $\langle T_{\th\th}\rangle$.

The parameter $Y$ has two solutions
for the different choices of the sign in eq.\eqref{Y-sol}.
In the $\kappa$-expansion,
the solution with the $-$ sign is approximately
\be
Y = \frac{1}{a^2} + \kappa \frac{6\b + 4\c}{a^4} 
+ \mathcal{O}(\kappa^2).
\label{Y1}
\ee
The solution with the $+$ sign is approximately
\be
Y = \frac{3}{8 \kappa \b} - \frac{4\b + 3\c}{2 a^2 \b} 
+ \mathcal{O}(\kappa^2),
\ee
which blows up in the limit $\kappa \rightarrow 0$.
This means that the energy-momentum tensor is not 
that of the vacuum of the conformal field.
The same comments below eq.\eqref{B3-sol2} apply here.

Demanding that the solution has a classical limit,
we should take the first solution \eqref{Y1} of $Y$.
Then
\be
B_2 = 1 + \kappa\frac{6\b + 4\c}{a^2}
- \kappa \langle T_{\th\th}(z_0)\rangle + \mathcal{O}(\kappa^2).
\ee
When the conformal charges $\b$, $\c$ vanish,
in the absence of the pressure $\langle T_{\th\th}(z_0)\rangle$,
this expression reproduces the result $B_2 = 1$ 
of the Schwarzschild solution.
(But there is no neck in the Schwarzschild solution at $z = z_0$
because it coincides with the horizon.)
For a large black hole,
since $\kappa$ is very small,
$B_2$ is positive unless
$\langle T_{\th\th}(z_0) \rangle \gtrsim \mathcal{O}(\kappa^{-1})$,
and there is a local minimum of the areal radius at $z = z_0$.

Next we can solve the coefficients $A_3$ and $B_3$.
They are
\begin{align}
A_3 &=
- 2\left[{3 a^2 \left(
- 3 a^2 + 8 \kappa\b + 16 \kappa a^2\b Y
- 24 \kappa^2\b \langle T_{\th\th}(z_0)\rangle + 12 \kappa\c\right)}\right]^{-1}
\nn \\
&
\left[
\frac{3}{8} a^4 A_1^3 
- 2\kappa a^4\b A_1^3 Y + 6\kappa a^4\c A_1 Y^2 
- \kappa a^2\b A_1^3 
- \frac{3}{2}\kappa a^2\c A_1^3
- \frac{3}{2}\kappa a^2 A_1 \langle T_{\th\th}(z_0)\rangle 
\right.
\nn \\
&
+ 3\kappa^2 a^2\b A_1^3 \langle T_{\th\th}(z_0)\rangle
+ 8\kappa^2 a^2\b A_1 \langle T_{\th\th}(z_0)\rangle Y 
- 12\kappa^2 a^2\c A_1 \langle T_{\th\th}(z_0)\rangle Y
+ 4\kappa^2\b A_1 \langle T_{\th\th}(z_0)\rangle 
\nn \\
&
- 6\kappa^2\b \langle T_{\th\th}'(z_0)\rangle 
+ 6\kappa^2\c A_1 \langle T_{\th\th}(z_0)\rangle 
- 12\kappa^2\c \langle T_{\th\th}'(z_0)\rangle
- 12\kappa^2 a^2\b Y \langle T_{\th\th}'(z_0)\rangle
\nn \\
&
\left.
- 12 \kappa^3\b A_1 \langle T_{\th\th}(z_0)\rangle^2
+ 6 \kappa^3\c A_1 \langle T_{\th\th}(z_0)\rangle^2
+ 18 \kappa^3\b \langle T_{\th\th}(z_0)\rangle \langle T_{\th\th}'(z_0)\rangle
\right]
\nn \\
&
\simeq \frac{1}{12} A_1^3 
+ \frac{\kappa}{24}A_1\left[32\c Y^2 
- \frac{\langle T_{\th\th}(z_0)\rangle}{a^2} \right]
+ \mathcal{O}(\kappa^2),
%\nn \\
%&
%\simeq \frac{1}{12} A_1^3 
%+ \frac{2\kappa}{3a^4}\left[2cA_1
%- \frac{1}{2} a^2 A_1 \langle T_{\th\th}(z_0)\rangle\right]
%+ \mathcal{O}(\kappa^2).
\displaybreak[1]\\
B_3 &=
- \left[3 \left(- 3a^2 + \kappa(8\b+12\c+16a^2\b Y) 
- 24\kappa^2\b\langle T_{\th\th}(z_0)\rangle \right)\right]^{-1}
\nn \\
&
\left[- \frac{3}{2} a^4 A_1 Y
+ 8\kappa a^4\b A_1 Y^2 + 12\kappa a^4\c A_1 Y^2
+ 4\kappa a^2\b A_1 Y
+ \frac{3}{2}\kappa a^2 A_1 \langle T_{\th\th}(z_0)\rangle
\right.
\nn \\
&
\left.
- 3\kappa a^2 \langle T_{\th\th}'(z_0)\rangle
+ 6\kappa a^2\c A_1 Y
\right.
\nn \\
&
- 24\kappa^2 a^2\c A_1 \langle T_{\th\th}(z_0)\rangle Y
- 20\kappa^2 a^2\b A_1 \langle T_{\th\th}(z_0)\rangle Y
- 4\kappa^2\b \langle T_{\th\th}'(z_0)\rangle
- 8\kappa^2 a^2\b Y \langle T_{\th\th}'(z_0)\rangle
\nn \\
&
- 6\kappa^2\c A_1 \langle T_{\th\th}(z_0)\rangle 
- 12\kappa^2\c \langle T_{\th\th}'(z_0)\rangle
\nn \\
&
\left.
+ \kappa^3 12\b A_1 \langle T_{\th\th}(z_0)\rangle^2 
- 4\kappa^2\b A_1 \langle T_{\th\th}(z_0)\rangle
+ 12\kappa^3\b \langle T_{\th\th}(z_0)\rangle \langle T_{\th\th}'(z_0)\rangle 
+ 12\kappa^3\c A_1 \langle T_{\th\th}(z_0)\rangle^2
\right]
\nn \\
%&
%\simeq - \frac{1}{6}a^2 A_1 Y 
%+ \frac{\kappa}{3}\left[
%4a^2 cA_1 Y^2 + \frac{1}{2} A_1 \langle T_{\th\th}(z_0)\rangle 
%- \langle T'_{\th\th}(z_0)\rangle
%\right]
%+ \mathcal{O}(\kappa^2)
%\nn \\
&\simeq - \frac{A_1}{6} 
- \frac{\kappa}{3} \left[
A_1(3\b-2\c) - \frac{1}{2} A_1 \langle T_{\th\th}(z_0)\rangle 
+ \langle T'_{\th\th}(z_0)\rangle\right]
+ \mathcal{O}(\kappa^2).
\end{align}
In the last line of the expressions of $A_3$ and $B_3$,
we have used the value of $Y$ in eq.\eqref{Y1}.

Plugging these back into $A(z)$ and $B(z)$, 
we get
\begin{align}
A(z) &= 
A_0 + A_1 z + \left(\frac{1}{a^2} - \frac{1}{4}A_1^2\right)z^2 
+ \frac{A_1^3}{12} z^3 + \mathcal{O}(z^4)
\nn \\
&
+ \kappa\left[
\frac{2(3\b+2\c)}{a^4}z^2
+ \frac{A_1\left(4\c - a^2 \langle T_{\th\th}(z_0)\rangle\right)}{3a^4}z^3
+ \mathcal{O}(z^4)
\right] + \mathcal{O}(\kappa^2),
\\
B(z) &=
a^2 + z^2 - \frac{A_1}{6} z^3 + \mathcal{O}(z^4)
\nn \\
&
+ \kappa\left[
\frac{2(3\b+2\c)}{a^2}z^2 
- \frac{2(3\b-2\c)A_1 - A_1 \langle T_{\th\th}(z_0)\rangle 
+ 2\langle T'_{\th\th}(z_0) \rangle}{6} z^3
+ \mathcal{O}(z^4)
\right] + \mathcal{O}(\kappa^2).
\end{align}

For given Weyl anomaly (given $\b$ and $\c$)
and given tangential pressure at the neck
$\langle T_{\th\th}(z_0)\rangle$,
these solutions are parametrized by 
3 parameters: $a$, $A_1$ and $z_0$.
However,
this solution around the neck should be continuously connected
to the asymptotic Schwarzschild solution at large distances.
We expect that once the classical Schwarzschild radius $a_0$ is fixed,
only two parametric constants remain independent,
corresponding to the freedom in specifying the parameter $q_{z_0}$
and the tangential pressure at the neck $\langle T_{\th\th}(z_0)\rangle$.

%{\color{blue}
%What are $\langle T^t{}_t \rangle$, $\langle T^z{}_z \rangle$
%and $\langle T^{\th}{}_{\th} \rangle$ at the neck?
%}

%%% pm %%%
%\subsection{Horizon ($q=0$)}
\subsection{Event Horizon ($q = 0$)}
%%%
\label{nonpert-horizon}

In this subsection,
we consider the class of solutions with even horizons. 
For $q=0$, the third term in \eqref{A-in-B} vanishes. 
As is discussed in Sec.~\ref{nonpert-wormhole}, for the expansion \eqref{B-exp-np}, 
$A(z)$ has the logarithmic divergence at $z=z_0$ and is expanded as 
\begin{equation}
 A(z) = A_0 + \frac{1 + a^2 \kappa f(z_0)}{B_2} \frac{1}{z-z_0} + \cdots. 
\end{equation}
This implies that there is the event horizon at $z=z_0$. 
The coefficients in the expansion of $B(z)$ are determined by \eqref{Eqtrace-nonpert} 
and calculated, for example, as 
\begin{align}
 B_2 &= \frac{1}{2} \left(1 + a^2 \kappa f(z_0)\right),  
\\
 B_3 &= \frac{a^2}{3} f'(z_0). 
\end{align}
Hence $A(z)$ and $B(z)$ are expanded as 
\begin{align}
 A(z) 
 &= 
 A_0 + 2 \log \left(z-z_0\right) + \mathcal O\left(\left(z-z_0\right)^2\right), 
\\
 B(z) 
 &= 
 a^2 + \frac{1}{2} \left(1 + a^2 \kappa f(z_0)\right) \left(z-z_0\right)^2 
 + \mathcal O\left(\left(z-z_0\right)^3\right). 
\end{align}
This is consistent with the results
\eqref{B-pert-z-horizon} and \eqref{B-x-horizon}
in the previous sections.

%This means that $A(z)$ diverges as $z$ approaches to the horizon at, say, $z_0$.
In a small neighborhood of the horizon,
we take the ansatz
\begin{align}
A(z) &= A_0 + 2d_0\log(z + d_1 z^2 + d_2 z^3 + d_3 z^4 + \cdots),
\\
B(z) &= a^2 + B_1 z + B_2 z^2 + B_3 z^3 + B_4 z^4 + \cdots,
\end{align}
where we have shifted the $z$-coordinate such that
the horizon is located at $z = 0$ and
we can also scale the $t$-coordinate to set 
$A_0 = 0$.

The semi-classical Einstein equations 
\eqref{Eqzz-nonpert}, \eqref{Eqtrace-nonpert}
are solved order by order in the $z$-expansion.
We find
\begin{align}
d_0 &= 1,
\\
B_1 &= 0,
\\
d_1 &= 0,
\\
B_2 &= 
\frac{a^2}{8\kappa(4\b+3\c)}\Big[
3-\frac{4\kappa(2\b+3\c)}{a^2}
+\frac{8\kappa^2 \b\langle T_{\th\th}(0)\rangle}{a^2}
\nn \\
&
\pm
\frac{\sqrt{3}}{a^2}
\Big(3a^4-48\kappa a^2(\b+\c)
+48\kappa^2 ((\b+\c)\c+a^2\b\langle T_{\th\th}(0)\rangle)
+24\kappa^2 a^2\c\langle T_{\th\th}(0)\rangle
\nn \\
&
+96\kappa^3 (\b+\c)c\langle T_{\th\th}(0)\rangle
-16\kappa^4\b\c\langle T_{\th\th}(0)\rangle^2
\Big)^{1/2}
\Big],
\\
d_2 &= \frac{4 B_2 + \kappa \langle T_{\th\th}(0)\rangle}{6 a^2}.
\end{align}

The choice of the plus sign in the solution of $B_2$ gives
\be
B_2 = \frac{1}{2} 
+ \kappa\left(\frac{3(\b+\c)}{a^2}-\frac{\langle T_{\th\th}\rangle}{2}\right)
+ \mathcal{O}(\kappa^2).
\ee
The choice of the minus sign gives a result
\be
B_2 = \frac{3a^2}{4\kappa(4\b+3\c)} - \frac{8\b+9\c}{8\b+6\c}
+ \mathcal{O}(\kappa)
\ee
that diverges as $\kappa\rightarrow 0$.
The latter case clearly does not correspond to a vacuum state.
The same comments below eq.\eqref{B3-sol2} apply here.

With $B_2$ given by the solution
with a finite value in the limit $\kappa\rightarrow\infty$,
we find
\begin{align}
B_3 &= - \frac{2\kappa}{9}\langle T'_{\th\th}(0)\rangle 
+ \mathcal{O}(\kappa^2),
\nn \\
d_3 &= - \frac{\kappa}{18a^2} \langle T'_{\th\th}(0)\rangle
+ \mathcal{O}(\kappa^2).
\end{align}

Since $B_1 = 0$ and $B_2 > 0$,
the function $B(z)$ also has a local extremum at $z = 0$.
But since $e^{A(z)}$ vanishes at $z = 0$,
it is a horizon and the solution does not apply to the range $z < 0$.
Hence it is in fact not really a local minimum.

For a given Weyl anomaly (given $\b$ and $\c$)
and a given $\langle T_{\th\th}(0)\rangle$,
these solutions are parametrized by 
the quantum Schwarzschild radius $a$
and the location of the horizon.
When the solution is identified as part of 
a global solution which asymptotes to
the Schwarzschild solution at large distances
with a given classical Schwarzschild radius,
there is only a single independent parameter,
corresponding to the choice of the tangential pressure
$\langle T_{\th\th}(0)\rangle$.

This is compatible with the perturbative analysis,
where it has been shown that the existence 
of the horizon demands a fine-tuning of a parameter $q_{z_0}$
to $q_{z_0} = 0$ exactly.
Hence the number of parameters 
for the solutions with a horizon
is one fewer than the other two classes of solutions.
%Our conclusion is that 
%the presence of a horizon demands fine tuning.

%{\color{blue}
%What are $\langle T^t{}_t \rangle$, $\langle T^z{}_z \rangle$
%and $\langle T^{\th}{}_{\th} \rangle$ at the horizon?
%}

%%% pm %%%
%\subsection{No Horizon or Neck ($q>0$)}
\subsection{No Neck \& No Horizon ($q > 0$)}
%%%

As we discussed in Sec.~\ref{nonpert-wormhole}, 
the expansion \eqref{B-exp-np} for $q>0$ gives 
complex $A(z)$ at $z=z_0$. 
Since $A(z)$ must be real, 
this implies that $r$ has no local minimum in this case.  
Hence $B_1$ must be non-zero for $q>0$, 
and $K(z)$ and $A(z)$ have regular expansions only with positive powers 
%%% pm %%%
of $z$
%%%
and thus there is no horizon.

\section{Conclusion}

We find in this paper that
the event horizon of the Schwarzschild solution
can be removed by the back reaction of quantum energy.
Depending on the quantum state,
more precisely,
depending on the parameter $q$
in $\langle T^r{}_r \rangle$ \eqref{Tupdownrr},
the near-horizon region can resemble a wormhole throat for $q < 0$,
or it has a horizon for $q = 0$,
or it has neither a throat nor a horizon for $q > 0$.
Notice that the horizon persists only if 
the parameter $q$ is fine-tuned to exactly zero.
The black hole is horizonless as long as $q$ is not exactly zero.
%%% pmh %%%
For definiteness,
we have focused on 4D conformal matter fields
so that the trace $\langle T^{\mu}{}_{\mu} \rangle$ is uniquely fixed.
But it should be clear from our calculation that,
for a wide class of models,
one would reach the same conclusion about 
the necessity of fine-tuning for horizon.
%while the parameter $q$ can always be defined by eq.\eqref{Eqzz-nonpert}.

It is interesting to ask whether the parameter $q$
is always driven to vanish for a generic gravitational collapse.
In the conventional model of black holes,
the answer is proposed to be affirmative.
But it is merely a folklore in the absence of 
a thorough study of gravitational collapses.
On the other hand,
in view of the fuzzball scenario \cite{fuzzball} 
or the firewall proposal \cite{firewall},
the answer could be negative.
A support for the negative answer also comes from 2D models.
For the 2D model of vacuum energy used in Ref.\cite{Davies:1976ei},
the wormhole-like structure is observed to appear 
in a generic gravitational collapse
in both analytical \cite{Ho:2018jkm}
and numerical studies \cite{Parentani:1994ij}.

Although we have only focused on static configurations in this paper.
The quantum vacuum state outside the collapsing matter 
for an astronomical black hole is expected to change very slowly over time.
The static solution should serve as a good approximation
of this time-dependent process.
It will therefore be very interesting to see how 
different models of evaporating black holes 
are connected to different solutions in this paper.
For instance,
a 4D self-consistent model which considers both the formation and evaporation processes
\cite{Kawai:2013mda,Kawai:2014afa,Ho:2015fja,Kawai:2015uya,Ho:2015vga,Ho:2016acf}
was shown to be compatible with 4D conformal anomaly \cite{Kawai:2017txu},
so it can be viewed as approximate time-dependent solutions
corresponding to some of the solutions in this paper.
As it does not have a horizon nor a wormhole-like throat,
but it has a Planck-scale tangential pressure,
the model is most likely related to the class of solutions
given in eqs.\eqref{A1-sol}, \eqref{A2-sol2}--\eqref{B3-sol2}.

%%%%%%%%%%%5
\section*{Acknowledgement}
We thank Chong-Sun Chu, Wen-Yu Wen, Chen-Pin Yeh, Shoichi Kawamoto, 
Po Ning and Philip Argyrus for discussions. 
P.M.H.\ and Y.M.\ are supported in part by the Ministry of Science and Technology, R.O.C. 
(project no.\ 104-2112-M-002-003-MY3) and by National Taiwan University (project no.\ 105R8700-2). 
H.K.\ is partially supported by Japan Society of Promotion of Science (JSPS), 
Grants-in-Aid for Scientific Research (KAKENHI) Grants No.16K05322.
Y.Y.\ is partially supported by Japan Society of Promotion of Science (JSPS), 
Grants-in-Aid for Scientific Research (KAKENHI) Grants No.\ 18K13550 and 17H01148. 
Y.Y.\ is also partially supported by RIKEN iTHEMS Program.

\hide{%%% beginning of hide %%%
\section*{Appendix}
Then, $A(z)$ is expressed in terms of $B(z)$ as 
\begin{equation}
 A(z) = \int^z F_2(z') dz' 
 + \log\left[A_0 - \int^z \left(e^{-\int^{z'} F_2(z'') dz''} F_1(z') dz'\right)\right], 
\end{equation}
where 
\begin{align}
 F_1(z) 
&= 
 \frac{2q B^2(z)}{B'(z)}, 
\\
 F_2(z) 
&= 
 \frac{2}{B'(z)} \left(1 + B(z) f(z)\right) + \frac{3B'(z)}{2B(z)}, 
\end{align}
and $A_0$ is the integration constant. 
If $B(z)$ and $f(z)$ have the following expansion around $z=0$; 
\begin{align}
 B(z) &= a^2 + B_2 z^2 + \cdots, 
\\
 f(z) &= f_0 + f_1 z + f_2 z^2 + \cdots, 
\end{align}
$A(z)$ is expanded as 
\begin{align}
 A(z) 
&= 
 \log\left(\frac{a^4 q}{1 + a^2 f_0}\right) - \frac{a^2 f_1 z}{1- a^2 f_0 - B_2} + \cdots 
 + A_0 \frac{1 + a^2 f_0 }{a^4 q} z^\frac{1+a^2 f_0}{B_2} + \cdots,  
\end{align}
if $q\neq 0$, and 
\begin{equation}
 A(z) = \frac{1+a^2 f_0}{B_2} \log z + \cdots, 
\end{equation}
for $q=0$. This implies that there is a horizon for $q=0$ but 
it becomes the local minimum of $r$ for $q\neq 0$. 
}%%% end of hide %%%

% example of figure and figure in the margin 
% can be found at the end of the file.

%\end{CJK} 
\end{document}